\documentclass[11pt]{article}
\usepackage{amssymb,color}
\newcommand{\N}{N\raise.7ex\hbox{\underline{$\circ $}}$\;$}

\begin{document}

\title{Electromagnetic Field in de Sitter
Expanding Universe: Majorana--Oppenheimer Formalism,\\ Exact
Solutions in non-Static Coordinates}

\author{
O.V. Veko\footnote{Kalinkovichi Gymnasium,
Belarus,vekoolga@mail.ru},
N.D Vlasii\footnote{N.N. Bogolyubov
Institute for Theoretical Physics. NAS of Ukraine},
Yu.A. Sitenko\footnote{N.N. Bogolyubov Institute for Theoretical
Physics. NAS of Ukraine, yusitenko@bitp.kiev.ua},
E.M. Ovsiyuk\footnote{Mozyr State Pedagogical University, e.ovsiyuk@mail.ru},
V.M. Red'kov\footnote{B.I. Stepanov Institute of Physics, NAS of
Belarus, redkov@dragon.bas-net.by}}

\maketitle

\begin{abstract}

Tetrad-based generalized complex  formalism by
Ma\-jorana--Oppenheimer is applied to treat electromagnetic field
in extending de Sitter Universe in non-static
spherically-symmetric coordinates. With the help of Wigner
$D$-functions, we separate angular dependence in the complex
vector field $E_{j}(t,r)+i B_{j}(t,r)$ from $(t,r)$-dependence.
The separation parameter arising here instead of frequency
$\omega$ in Minkowski space-time is quantized,
 non-static geometry of the de Sitter model leads to definite dependence of electromagnetic modes
 on the time variable.

Relation of 3-vector complex approach to 10-dimensional
Duffin--Kemmer--Petiau formalism is considered. On this base, the
electromagnetic waves of magnetic and electric type have been
constructed in both approaches. In Duffin--Kemmer--Petiau
approach, there are constructed gradient-type solutions in Lorentz
gauge.

\end{abstract}



Kkeywords{de Sitter Universe, electromagnetic field,
Majorana--Oppenheimer formalism, non-static coordinates, electric
and magnetic waves}

\section{Introduction}

Special relativity arose from study of  symmetry properties of
the Maxwell equations with respect to a motion of a reference frame:
Lorentz  \cite{1904-Lorentz},  Poincar\'{e} \cite{1905-Poincare, 1905-Poincare(2)},
Einstein \cite{1905-Einstein}. Indeed, the analysis of the
Maxwell equations with respect to the Lorentz transformations was the
first object of the special relativity:
 Minkowski \cite{1908-Minkowski},
 Silberstein \cite{1907-Silberstein(1), 1907-Silberstein(2), 1914-Silberstein(3)},
Marcolongo \cite{1914-Marcolongo},  Bateman  \cite{1915-Bateman},
 Lanczos \cite{1919-Lanczos}, Gordon \cite{1923-Gordon},
Mandel'stam --  Tamm
\cite{1925-Mandel'stam, 1925-Tamm(1), 1925-Tamm(2)}.
After discovery of the relativistic equation for a particle with
spin 1/2 -- Dirac \cite{1928-Dirac, 1928-Dirac(2)} -- much work was done to study
spinor and vectors in the context of the Lorentz group theory: M\"{o}glich
\cite{1928-Moglich},  Ivanenko -- Landau
\cite{1928-Ivanenko-Landau}, Neumann \cite{1929-Neumann}, van der
Waerden \cite{1929-Waerden}, Juvet \cite{1930-Juvet}. As was shown,
any quantity which transforms linearly under the Lorentz
transformations is a spinor. For that reason spinor quantities are
considered as fundamental in quantum field theory and basic
equations in particle physics should be written in a spinor form.
 For the first time, spinor formulation of the Maxwell equations was studied by Laporte
and Uhlenbeck \cite{1931-Laporte}, also see Rumer
\cite{1936-Rumer}. In 1931,   Majorana \cite{1931-Majorana} and
Oppenheimer \cite{1931-Oppenheimer} proposed to consider the
Maxwell theory of electromagnetism as a basis for wave mechanics of the
photon. They introduced a  complex 3-vector wave function
satisfying the massless Dirac-like equations. Before Majorana  and
Oppenheimer,    the most crucial steps were made by Silberstein
\cite{1907-Silberstein(1)}, he showed the possibility to consider
 Maxwell equations in terms of complex 3-vector variables.
In his second paper Silberstein   \cite{1907-Silberstein(2)} noted
that the complex form of the Maxwell equations had  been already known;
he referred to the lecture notes on
differential equations of mathematical physics given by B. Riemann that
were edited and published by H. Weber in 1901  \cite{1901-Weber}.
 This is not widely used fact as  noted by Bialynicki-Birula   \cite{1994-Bialynicki-Birula, 1996-Bialynicki-Birula}).

And several general remarks. Maxwell equations in various  forms were  considered
during long time by many authors. In this paper we did not discuss  in detail did not  classify
 these contributions -- it should be a matter of special consideration. Instead we just try
  to give an historically organized  list.
   They are:  Luis de Broglie
\cite{1934-Broglie(1), 1934-Broglie(2), 1939-Broglie, 1940-Broglie},
Mercier \cite{1935-Mercier},
 Petiau \cite{1936-Petiau}, Proca \cite{1936-Proca, 1946-Proca},
Duffin \cite{1938-Duffin}, Kemmer
\cite{1939-Kemmer, 1943-Kemmer, 1960-Kemmer},  Bhabha
\cite{1939-Bhabha}, Belinfante
\cite{1939-Belinfante(1), 1939-Belinfante(2)}, Taub
\cite{1939-Taub},  Sakata  -- Taketani \cite{1940-Sakata},
Schr\"{o}dinger  \cite{1940-Schrodinger, 1943-Schrodinger(1), 1943-Schrodinger(2)}, Tonnelat
\cite{1941-Tonnelat}, Heitler \cite{1943-Heitler},
\cite{1946-Harish-Chandra(1), 1946-Harish-Chandra(2)},
Hoffmann \cite{1947-Hoffmann}, Utiyama \cite{1947-Utiyama},
Mercier \cite{1949-Mercier},   Fujiwara
\cite{1953-Fujiwara}, G\"{u}rsey \cite{1954-Gursey}, Gupta
\cite{1954-Gupta}, Lichnerowicz \cite{1954-Lichnerowicz},  Ohmura
\cite{1956-Ohmura},
  Borgardt \cite{1956-Borgardt, 1958-Borgardt},  Fedorov \cite{1957-Fedorov},  Kuohsien \cite{1957-Kuohsien},
Bludman   \cite{1957-Bludman}, Good \cite{1957-Good},  Moses
\cite{1958-Moses, 1959-Moses},  Silveira
\cite{1959-Silveira},
Lomont
\cite{1958-Lomont},  Post
\cite{1961-Post}, Bogush -- Fedorov \cite{1962-Bogush-Fedorov},
Sachs -- Schwebel  \cite{1962-Sachs-Schwebel},  Newman -- Penrose \cite{Newman-Penrose}
 Ellis
\cite{1964-Ellis}, Oliver \cite{1968-Oliver}, Beckers -- Pirotte
\cite{1968-Beckers},  Casanova \cite{1969-Casanova},  Carmeli
\cite{1969-Carmeli},
Bogush \cite{1971-Bogush}, Goldman -- Tsai -- Yildiz \cite{Goldman-Tsai-Yildiz},
Lord
\cite{1972-Lord}, Weingarten \cite{1973-Weingarten},   Mignani --
Recami --  Baldo  \cite{1974-Recami1}, \cite{1974-Frankel},
Stephani {Stephani},
Edmonds \cite{1975-Edmonds},  Strazhev --
Tomil'chik  \cite{1975-Strazhev-Tomil'chik},
Fushchych -- Nikitin \cite{Fushchych-Nikitin},
Malin \cite{Malin}, Silveira
\cite{1980-Silveira},
Frolov {Frolov},  Jena -- Naik --  Pradhan \cite{1980-Jena},
Venuri  \cite{1981-Venuri}, Chow \cite{1981-Chow}, Fushchich --
Vladimirov  \cite{Fushchych-Vladimirov}, Fushchich -- Nikitin \cite{1983-Fushchich, Fushchich-Nikitin},
 Cook \cite{1982-Cook(1), 1982-Cook(2)},
  Giannetto \cite{1985-Giannetto},
Recami \cite{1990-Recami},
Figueiredo,  Oliveira and  Rodrigues  \cite{Figueiredo-Oliveira-Rodrigues(1990)},
 Krivsky --  Simulik \cite{1992-Krivsky},
Vaz and Rodrigues \cite{1993-Vaz-Rodrigues, 1998-odrigues-Vaz},
 Inagaki \cite{1994-Inagaki},
Bialynicki-Birula \cite{1994-Bialynicki-Birula, 1996-Bialynicki-Birula,  2005-Birula},
Sipe \cite{1995-Sipe}, \cite{1996-Ghose},
Dvoeglazov  \cite{1998-Dvoeglazov', 1998-Dvoeglazov, 2001-Dvoeglazov},
Gersten  \cite{1998-Gersten},
Gsponer \cite{2002-Gsponer},
Varlamov  \cite{2002-Varlamov},
Khan \cite{2002-Khan(1),  2002-Khan(2)},
Donev --  Tashkova \cite{2004-Donev(1), 2004-Donev(2), 2004-Donev(3)},
Rodrigues et al \cite{Figueiredo-Oliveira-Rodrigues(1990),1993-Vaz-Rodrigues,
1998-odrigues-Vaz}, Bogush et al
\cite{Bogush-Kisel-Tokarevskaya-Red'kov},
Ovsiyuk and Red'kov \cite{Ovsiyuk-Red'kov, Book-2009,
Ovsiyuk-Red'kov-2011'}.

The interest in the
Majorana-Oppenheimer formulation of electrodynamics has grown in
recent years.
In particular,
elaboration  of the complex formalism in electrodynamics to Riemannian space-time models
 has been performed in recent years -- see
\cite{Red'kov-Tolkachev-2010,
Bogush-Krylov-Ovsiyuk-Red'kov-2010, Red'kov-Ovsiyuk-2011,
Kisel-Ovsiyuk-Red'kov-Tokarevskaya-2011,
 Red'kov-Tolkachev-2012,
Red'kov-Tokarevskaya-Spix-2012, Red'kov-Tokarevskaya-Spix-2013,
Ovsiyuk-Red'kov-Tokarevskaya-2013}.

In general, study  fundamental  particle fields on the background of expanding universe,
in particular de Sitter and anti de Sitter models,
has a long history    [145-171]. Special value of these  geometries consists in their simplicity and high symmetry
groups underlying them
which makes us to believe in existence of exact analytical treatment for  some fundamental problems of
classical and quantum field theory  in curved spaces. In particular, there exist special representations for
fundamental wave equations, Dirac's and Maxwell's, which are explicitly invariant under  respective
symmetry groups
$SO(4.1)$  and $SO(3.2)$ for these models.

In the present paper, the Ma\-jorana--Oppenheimer approach and Duffin--Kemmer--Petiau formalism
will be applied to treat electromagnetic field
in extending de Sitter Universe in non-static
spherically-symmetric coordinates. With the help of Wigner
$D$-functions, we separate angular dependence from $(t,r)$-variables. The
separation parameter arising here instead of frequency $\omega$ in
Minkowski space-time is quantized,
 non-static geometry of the de Sitter model leads to definite dependence of electromagnetic modes
 on the time variable.
Relation of 3-vector complex approach to 10-dimensional
Duffin--Kemmer--Petiau formalism is elaborated. On this base, the
electromagnetic waves of magnetic and electric type have been
constructed in both approaches. Besides, in Duffin--Kemmer--Petiau
approach, there are constructed gradient-type solutions.

The popular group theoretical approach for spin one field (massive and massless cases)
 will be considered for treating the problem os spin one field in extending de Sitter Universe in a separate paper.

\section{Electromagnetic field in Majorana--Oppenheimer formalism}

The matrix form of Maxwell equations in Majorana--Oppenheimer
formalism is \cite{Book-2009}
\begin{eqnarray}
 \alpha^{c} \; ( \; e_{(c)}^{\rho} \partial_{\rho}  + {1
\over 2} j^{ab} \gamma_{abc} \; ) \; \Psi =0\; ,
\nonumber
\\
 \alpha^{0} = -i I\; ,
\quad \Psi = \left | \begin{array}{c} 0 \\ {\bf E} + i c{\bf B}
\end{array} \right |  \; ,
\label{A.1}
\end{eqnarray}

\noindent or in a more detailed form
\begin{eqnarray}
-i  \; ( \; e_{(0)}^{\rho} \partial_{\rho}  +  {1 \over 2} j^{ab}
\gamma_{ab0} \; )\Psi
\nonumber\\
 + \alpha^{k} \; ( \; e_{(k)}^{\rho}
\partial_{\rho}  + {1 \over 2} j^{ab} \gamma_{abk} \; )\Psi =0
\; .
\label{A.2}
\end{eqnarray}

We will need  expressions for matrices $\alpha^{k}$ and six
generators of the 3-vector complex representation of the group
$SO(3,C)$ (first specify them in Cartesian basis)
\begin{eqnarray}
\alpha^{1} = \left | \begin{array}{rrrr}
0 & 1  &  0  & 0  \\
-1 & 0  &  0  & 0  \\
0 & 0  &  0  & -1  \\
0 & 0  &  1  & 0
\end{array}  \right |,
\;\;
\alpha^{2} = \left | \begin{array}{rrrr}
0 & 0  &  1  & 0  \\
0 & 0  &  0  & 1  \\
-1 & 0  &  0  & 0  \\
0 & -1  & 0  & 0
\end{array}  \right |,
\;\;
\alpha^{3} = \left | \begin{array}{rrrr}
0 & 0  &  0  & 1  \\
0 & 0  & -1  & 0  \\
0 & 1  &  0  & 0  \\
-1 & 0  &  0  & 0
\end{array}  \right |  ,
\nonumber
\end{eqnarray}
\begin{eqnarray}
S^{1}= j^{23} =  \left | \begin{array}{cc} 0 & 0 \\ 0 & \tau_{1}
\end{array} \right |,
\;\; N^{2} = j^{01} =  +i \left | \begin{array}{cc} 0 & 0 \\ 0 &
\tau_{1}
\end{array} \right |,
\nonumber
\\
S^{2} = j^{31} =
 \left | \begin{array}{cc}
0 & 0 \\ 0 & \tau_{2}
\end{array} \right |  , \;\;
N^{2} =  j^{02} = +i \left | \begin{array}{cc} 0 & 0 \\ 0 &
\tau_{2}
\end{array} \right |   ,
\nonumber
\\
S^{3} = j^{12} = \left | \begin{array}{cc} 0 & 0 \\ 0 & \tau_{3}
\end{array} \right |,
\;\; N^{3} = j^{03} = +i
 \left | \begin{array}{cc}
0 & 0 \\ 0 & \tau_{3}
\end{array} \right |  ;
\nonumber
\end{eqnarray}

\noindent generators obey the commutation relations
\begin{eqnarray}
S^{1}  S^{2}  -S^{2} S^{1}   = S^{3}  , \nonumber \\ N^{1}  N^{2}
-N^{2} N^{1}  = -S^{3}  ,
\nonumber\\
 S^{1}N^{2} - N^{2}S^{1}  = +N^{3} ;
\nonumber
\end{eqnarray}

\noindent and similar ones in accordance with cyclic symmetry.
Below we will use notation
\begin{eqnarray}
j^{23} = S^{1}, \quad j^{31} = S^{2}, \quad j^{12} = S^{3}  ,
\nonumber\\[2mm]
j^{01} = i S^{1}, \quad j^{02} = iS^{2}, \quad j^{03} = iS^{3}
.
\nonumber
\end{eqnarray}

Let us consider eq.  (\ref{A.1}) in non-static coordinates of the Sitter space-time
\begin{eqnarray}
 x^{\alpha } =(x^{0}, x^{1}, x^{2}, x^{3}) = (t, r, \theta, \phi) \, ,
\nonumber
\\
 dS^{2} =  dt^{2} -  \cosh^{2} t  [
dr^{2} + \sin^{2} r ( d\theta ^{2} + \sin^{2} \theta  d\phi^{2} )
]
 \nonumber
\end{eqnarray}

\noindent at the use of the following tetrad
\begin{eqnarray}
e^{\alpha}_{(0)} = ( 1 , 0 , 0 , 0 ) \, , \quad e^{\alpha }_{(1)}
= ( 0 , 0 ,{ 1 \over   \cosh t \sin r}, 0 )\,  ,
\nonumber
\\
e^{\alpha }_{(2)} = ( 0 ,0 , 0 , {1 \over \cosh t \, \sin r \,  \sin
\theta})\, ,
\quad
 e^{\alpha }_{(3)} = ( 0 , { 1 \over   \cosh t
} , 0 , 0 ) \, . \label{A.6}
\end{eqnarray}
Non-vanishing Ricci rotation symbols are
\begin{eqnarray}
\gamma_{[01]1}= \tanh t \, ,  \quad \gamma_{[02]2}= \tanh t \, ,
\nonumber
\\ \gamma_{[03]3}= \tanh t \,  ,
\quad
\gamma_{[31]1}=  +{\cot r \over \cosh t} \, ,
\nonumber
\\
\gamma_{[32]2}= + {\cot r \over \cosh t} \, , \quad
\gamma_{[12]2} = {1 \over \cosh t}{ \cot \theta \over \sin r  }.
\nonumber
\end{eqnarray}

\noindent These permit us to find
 $1 / 2\, j^{ab} \gamma_{abk}$:
\begin{eqnarray}
{1 \over 2} j^{ab} \gamma_{ab1} = [ S^{1} ( \gamma_{231} +i
\gamma_{011} ) + S^{2} ( \gamma_{311} +i \gamma_{021})
\nonumber\\
 + S^{3} (
\gamma_{121} +i \gamma_{031} )\;  ]  =
  i S^{1}
 \tanh t   + S^{2} {\cot r \over \cosh t}\,,
\nonumber
\end{eqnarray}
\begin{eqnarray}
{1 \over 2} j^{ab} \gamma_{ab2} = [ S^{1} ( \gamma_{232} +i
\gamma_{012} ) + S^{2} ( \gamma_{312} +i \gamma_{022})
\nonumber\\
 + S^{3} (
\gamma_{122} +i \gamma_{032} )\,  ]
\nonumber\\
=- S^{1}  {\cot r \over \cosh t}
 + i S^{2}  \tanh t + s^{3}
{1 \over \cosh t}{ \cot \theta \over \sin r  }  \,,  \nonumber
\end{eqnarray}
\begin{eqnarray}
{1 \over 2} j^{ab} \gamma_{ab3} = [ S^{1} ( \gamma_{233} +i
\gamma_{013} ) + S^{2} ( \gamma_{313} +i \gamma_{023})
\nonumber\\
+ S^{3} (
\gamma_{123} +i \gamma_{033} )\;  ]
=  i S^{3}
  \tanh t \,.
\nonumber
\end{eqnarray}

\noindent
 Therefore, the matrix Maxwell equation
 \begin{eqnarray}
-i  \, ( \, \partial_{t}  +  {1 \over 2} j^{ab} \gamma_{ab0} \,
)\Psi
+ \alpha^{1} \, ( \, e_{(1)}^{2} \partial_{2}  + {1 \over 2}
j^{ab} \gamma_{ab1} \, )\Psi
\nonumber\\
+ \alpha^{2} \, ( \,e_{(2)}^{3}
\partial_{3}  + {1 \over 2} j^{ab} \gamma_{ab2} \, )\Psi
 +
\alpha^{3} \, ( \, e_{(3)}^{1} \partial_{1}  + {1 \over 2} j^{ab}
\gamma_{ab3} \, )\Psi=0
\nonumber
\end{eqnarray}

 \noindent reads
\begin{eqnarray}
-i  \,  \partial_{t}   \,\Psi + \alpha^{3} \,\left  ( \, { 1 \over
\cosh t}\, \partial_{r}  + {1 \over 2} j^{ab} \gamma_{ab3} \,
\right )\Psi
\nonumber\\
+ \alpha^{1} \,\left ( \, { 1 \over   \cosh t \sin r}\,
\partial_{\theta}  + {1 \over 2} j^{ab} \gamma_{ab1} \, \right )\Psi
+ \alpha^{2} \, \left ( \, {1 \over \cosh t  \sin r \,  \sin
\theta}\,
\partial_{\phi}  + {1 \over 2} j^{ab} \gamma_{ab2} \, \right )\Psi =0\,,
\nonumber
\end{eqnarray}

\noindent and then
\begin{eqnarray}
\left[-i  \,  \partial_{t} + \alpha^{3}  \left( \, { 1 \over
\cosh t}\, \partial_{r}  + i S^{3} \tanh t \,\right ) \right.
+ \alpha^{1}
\left(  { 1 \over   \cosh t \sin r}\, \partial_{\theta}  + i
S^{1}
 \tanh t   + S^{2} {\cot r \over \cosh t}  \right)
\nonumber
\\
+ \alpha^{2}  \left( \, {1 \over \cosh t  \sin r \,  \sin
\theta}\, \partial_{\phi}   - S^{1}  {\cot r \over \cosh t}\right.
 \left.\left.+ i S^{2}  \tanh t + S^{3}
{1 \over \cosh t}{ \cot \theta \over \sin r  }  \,\right
)\right]\Psi=0\,. \label{A.10}
\end{eqnarray}

Let us compare it with the structure of DKP 10-component (massive
case) equation for spin 1 field (see \cite{Book-2011}) in the same
coordinates and tetrad
\begin{eqnarray}
\left [   ( i\beta^{0}  {  \partial \over \partial t} -  m)   + i
\tanh t   \left ( \beta^{1} j^{01}    + \beta^{2} j^{02} +
\beta^{3} j^{03}   \right ) \right.
\nonumber
\\
  \left.     +{ i \over \cosh t}
   \left (  \beta^{3}  \partial_{r} +   {\beta^{1} j^{31}  +   \beta^{2} j^{32} \over \tan r}\right )  +
{1 \over \cosh t \sin r} \left (  i\beta^{1}   {\partial \over
\partial \theta} + \beta^{2} {  i \partial_{\phi}   +
    ij^{12}   \cos \theta \over  \sin \theta  }  \right ) \right  ]  \Psi = 0 \, .
\label{A.11a}
\end{eqnarray}

\noindent Performing in  (\ref{A.11a})  formal changes
\begin{eqnarray}
 i \beta^{0}  \;\; \Longrightarrow   \;\; -i , \;   i \beta^{k}
\;\; \Longrightarrow   \;\; \alpha^{k},
\qquad
 j^{0k}  \;\;
\Longrightarrow   \;\; iS^{k} ,
\;
j^{31} \;\; \Longrightarrow   \;\;  S^{2},
\qquad
 j^{32} \;\;
\Longrightarrow   \;\; - S^{1}, \quad j^{12} \;\; \Longrightarrow
\;\;  S^{3},
\nonumber
\end{eqnarray}

\noindent we get
\begin{eqnarray}
\left [    -i   {  \partial \over \partial t} -  m   +i
 \tanh t   \left ( \alpha^{1} S^{1}    + \alpha^{2} S^{2}
+ \alpha^{3} S^{3}   \right )
      +{ 1 \over \cosh t}
   \left (  \alpha^{3}  \partial_{r} +   {\alpha^{1} S^{2}  -   \alpha^{2} S^{1} \over \tan r}\right ) \right.
   \nonumber
\\
\left.    +
{1 \over \cosh t \sin r} \;  \alpha^{1}   {\partial \over
\partial \theta}
+ {1 \over \cosh t \sin r} \;\alpha^{2} {   \partial_{\phi}   +
    S^{3}   \cos \theta \over  \sin \theta  }   \right  ]  \Psi = 0  .
\label{A.11c}
\end{eqnarray}

\noindent Eq. (\ref{A.11c}) at  $m=0$ should coincide with
(\ref{A.10}) -- it is indeed so. Bellow, we will represent eq.
(\ref{A.10}) in the form closed to DKP-form:
\begin{eqnarray}
\left\{       -i   {  \partial \over \partial t}  +i
 \tanh t    ( \alpha^{1} S^{1}    + \alpha^{2} S^{2}
+ \alpha^{3} S^{3}    )
       +{ 1 \over \cosh t}
   \left (  \alpha^{3}  \partial_{r} +   {\alpha^{1} S^{2}  -   \alpha^{2} S^{1} \over \tan r}\right ) \right.
   \nonumber
\\
   \left. +
{1 \over \cosh t \sin r} \Sigma_{\theta\phi}  \right  \}  \; \Psi
= 0 \; ,\qquad
\Sigma_{\theta\phi} = \left (  \alpha^{1}   {\partial \over
\partial \theta} + \alpha^{2} {   \partial_{\phi}   +
    S^{3}   \cos \theta \over  \sin \theta  }  \right )\,.
\label{A.12}
\end{eqnarray}

To have the matrix  $j^{12}$ diagonal, we translate all description
to the cyclic basis:
$$
 \Psi ' = U_{4} \Psi  \, , \quad
 U_{4}  = \left | \begin{array}{cc} 1 & 0 \\
0 &   U
\end{array} \right |  ,
U = \left |
 \begin{array}{ccc}
- 1 /\sqrt{2}  &  i /  \sqrt{2}  &  0  \\[2mm]
0  &  0  &  1  \\[2mm]
1 / \sqrt{2}  &  i  / \sqrt{2}  &  0
\end{array} \right |  ,
U^{-1}  = U^{+}_{3} = \left |
 \begin{array}{ccc}
- 1 /\sqrt{2}  &  0  & 1 /  \sqrt{2}    \\[2mm]
-i / \sqrt{2}  &  0  &  -i / \sqrt{2}   \\[2mm]
0    &  1    &  0
\end{array} \right |  ;
$$
$$
U  \tau_{1} U^{-1} = {1 \over \sqrt{2}} \left |
\begin{array}{ccc}
0  &  -i   &  0  \\
-i  &  0  &  -i  \\
0  &  -i  &  0
\end{array} \right | =  \tau'_{1}  ,\qquad
j^{'23} = s_{1}' = \left | \begin{array}{cc}
0 & 0 \\
0 & \tau'_{1}
\end{array} \right |  ,
$$
$$
U  \tau_{2} U^{-1} = {1 \over \sqrt{2}} \left |
\begin{array}{ccc}
0  &  -1  &  0  \\
1  & 0  &  -1  \\
0  &  1  &  0
\end{array} \right | =  \tau'_{2} ,\qquad  j^{'31} = s_{2}' = \left | \begin{array}{cc}
0 & 0 \\
0 & \tau'_{2}
\end{array} \right |
,
$$
$$
U  \tau_{3} U^{-1} = - i \; \left | \begin{array}{rrr}
+1  &  0  &  0  \\
0  &  0  &  0   \\
0  &  0  &  -1
\end{array} \right | = \tau'_{3} \,,
\qquad
  j^{'12} = s_{3}' = \left | \begin{array}{cc}
0 & 0 \\
0 & \tau'_{3}
\end{array} \right |
;
$$
$$
\alpha^{'1} = {1 \over \sqrt{2}} \left | \begin{array}{rrrr}
0 & - 1 & 0 & 1 \\
1 & 0 & -i & 0 \\
0 & -i & 0 & - i \\
-1 & 0 & -i & 0
\end{array} \right |  ,
 \alpha^{'2} = {1 \over \sqrt{2}} \left | \begin{array}{rrrr}
0 & - i & 0 & -i \\
-i & 0 & -1 & 0 \\
0 & 1 & 0 &  -1 \\
-i & 0 & 1 & 0
\end{array} \right |  ,
 \alpha^{'3}  =
\left | \begin{array}{rrrr}
0  & 0  & 1 & 0 \\
0  & -i & 0 & 0 \\
-1 &  0 & 0 & 0 \\
0  &  0 & 0 & +i
\end{array} \right |  .
$$

 Equation (\ref{A.12})  formally preserves its form
 \begin{eqnarray}
\left\{       -i   {  \partial \over \partial t}  +i
 \tanh t    ( \alpha^{'1} S^{'1}    + \alpha^{2} S^{'2}
+ \alpha^{'3} S^{'3}    )  \right.
\nonumber
\\
       +{ 1 \over \cosh t}
   \left (  \alpha^{'3}  \partial_{r} +   {\alpha^{'1} S^{'2}  -   \alpha^{'2} S^{'1} \over \tan r}\right )
   \left. +
{1 \over \cosh t \sin r} \Sigma'_{\theta\phi}  \right  \}  \;
\Psi' = 0 \; ,
\nonumber
\\
\Sigma'_{\theta\phi} = \left (  \alpha^{'1}   {\partial \over
\partial \theta} + \alpha^{'2} {   \partial_{\phi}   +
    S^{'3}   \cos \theta \over  \sin \theta  }  \right )\,.
\label{A.14}
\end{eqnarray}

\noindent In the following, the primes will be omitted.

We will diagonalize the square and third projection of the total
angular  moment, corresponding substitution for field function is
 \cite{Book-2011}
\begin{eqnarray}
\psi  =  \left | \begin{array}{c}
0 \\
\varphi_{1}(t,r) D_{-1 }
\\
\varphi_{2}(t,r) D_{0 }  \\
\varphi_{3}(t,r) D_{+1 }
\end{array} \right |\,,
\label{A.15}
\end{eqnarray}

\noindent where Wigner $D$-functions \cite{VMH} are designated as
 $D_{\sigma} = D^{j}_{-m, \sigma} ( \phi , \theta, 0)$, $\sigma = -1, \,0,\, +1$;
 $j,\,m$ stand for angular moment quantum numbers. With the help of recurrent formulas
 \cite{VMH}
\begin{eqnarray}
\partial_{\theta} \,  D_{-1} =   {1 \over 2} \, ( \,
a  \, D_{-2} -    \nu  \, D_{0} \, ) \, ,
\nonumber
\\
 \frac {m -
\cos{\theta}}{\sin {\theta}} \, D_{-1} = {1 \over 2} \, ( \, a \,
D_{-2} + \nu \, D_{0} \, ) \, ,
\nonumber
\\
\partial_{\theta}  \, D_{0} = {1 \over 2} \, ( \,
\nu  \, D_{-1} - \nu  \, D_{+1} \, ) \, ,
\nonumber
\\
 \frac {m}{\sin
{\theta}} \, D_{0} = {1 \over 2} \, (  \,  \nu \, D_{-1} +  \nu \,
D_{+1} \, ) \, ,
\nonumber
\\
\partial_{\theta} \, D_{+1} =
{1 \over 2} \, ( \, \nu  \, D_{0} - a  \, D_{+2} \, ) \, ,
\nonumber
\\
\frac {m + \cos{\theta}}{\sin {\theta}}  \, D_{+1} ={1 \over 2} \,
( \,
 \nu \, D_{0} + \ a  \, D_{+2} \, ) \,  ,
\nonumber
\end{eqnarray}
where $$ \nu =  \sqrt{j(j+1)} \, , \;\; a = \sqrt{(j-1)(j+2)}\;,
$$

\noindent we find action of the angular operator
\begin{eqnarray}
 \Sigma_{\theta \phi} \Psi  =
 { \nu  \over \sqrt {2}} \;
 \left | \begin{array}{r}
 (\varphi_{1} +\varphi_{3}) D_{0} \\
 -i \;  \varphi_{2} D_{-1}
 \\
 i \;  (\varphi_{1} -\varphi_{3}) D_{0}
 \\
 + i \;  \varphi_{2} D_{+1}
 \end{array} \right  |\,.
\label{A.16b}
\end{eqnarray}

\noindent With the use of intermediate formulas

\begin{eqnarray}
-i   {  \partial \over \partial t}\,\Psi =-i\left |
\begin{array}{c}
0 \\
  \partial_{t} \varphi_{1} \;D_{-1 }
\\
  \partial _{t}\varphi_{2}\;D_{0 }  \\
\partial _{t} \varphi_{3}\;  D_{+1 }
\end{array} \right |\,,
i
 \tanh t   \; ( \alpha^{1} S^{1}    + \alpha^{2} S^{2}
+ \alpha^{3} S^{3}    )\,\Psi
= -2 i
 \tanh t \,
 \left | \begin{array}{c}
0 \\
\varphi_{1} D_{-1 }
\\
\varphi_{2} D_{0 }  \\
\varphi_{3} D_{+1 }
\end{array} \right |,
\nonumber
\\[4mm]
{ 1 \over \cosh t}
    \alpha^{3}  \partial_{r}\,\Psi=
{ 1 \over \cosh t}\left | \begin{array}{c}
\partial_{r}  \varphi_{2} D_{0 } \\
 -i\,\partial_{r} \varphi_{1} D_{-1 }
\\
0  \\
i\, \partial_{r}\varphi_{3} D_{+1 }
\end{array} \right |,
     {\alpha^{1} S^{2}  -   \alpha^{2} S^{1}
      \over \cosh t\;\tan r}\,\Psi
      ={1\over \cosh t\;\tan r}\left | \begin{array}{c}
2\varphi_{2} D_{0 }  \\
-i\,\varphi_{1} D_{-1 }
\\
0 \\
i\,\varphi_{3} D_{+1 }
\end{array} \right |
\label{A.17c}
\end{eqnarray}

\noindent the Maxwell equation (\ref{A.14}) takes the form

\begin{eqnarray}
-i\left | \begin{array}{c}
0 \\
  \partial _{t} \varphi_{1}\;  D_{-1 }
\\
  \partial _{t} \varphi_{2}\;  D_{0 }  \\
  \partial_{t} \varphi_{3}\;  D_{+1 }
\end{array} \right |+i
 \tanh t \,\left | \begin{array}{c}
0 \\
-\,\varphi_{1} D_{-1 }-i\,\varphi_{1} D_{-1 }
\\
-2\varphi_{2} D_{0 }  \\
 -\,\varphi_{3} D_{+1 }-i\,\varphi_{3} D_{+1 }
\end{array} \right |+{ 1 \over \cosh t}\left | \begin{array}{c}
  \partial_{r}  f_{2} D_{0 } \\
 -i\, \partial_{r}  f_{1} D_{-1 }
\\
0  \\
i\,   \partial_{r} f_{3} D_{+1 }
\end{array} \right |
\nonumber
\end{eqnarray}
\begin{eqnarray}
+{1\over \cosh t\;\tan r}\left | \begin{array}{c}
2\varphi_{2} D_{0 }  \\
-i\,\varphi_{1} D_{-1 }
\\
0 \\
i\,\varphi_{3} D_{+1 }
\end{array} \right |+
{ \nu  \over \sqrt {2}} \;{1 \over \cosh t \sin r}
 \left | \begin{array}{r}
 (\varphi_{1} +\varphi_{3}) D_{0} \\
 -i \;  \varphi_{2} D_{-1}
 \\
 i \;  (\varphi_{1} -\varphi_{3}) D_{0}
 \\
 + i \;  \varphi_{2} D_{+1}
 \end{array} \right  |=0\,.
\label{A.18}
\end{eqnarray}

\noindent From whence it follows the system
\begin{eqnarray}
 \left ( {  \partial \over \partial r}+{2\over \tan r} \right ) \varphi_{2}+
 {\nu / \sqrt{2}  \over \sin r} (\varphi_{1} + \varphi_{3})=0\,,
\nonumber
\\
- \left ( \cosh t  {  \partial \over \partial t}+ 2
 \sinh t  \right ) \, \varphi_{1}
 - \left ( {  \partial \over \partial r} +{1\over \tan r} \right ) \varphi_{1}-
 {\nu /\sqrt{2} \over  \sin r} \,\varphi_{2} =0\,,
\nonumber
\\
- \left ( \cosh t {  \partial \over \partial t} + 2\,
 \sinh t \right ) \varphi_{2}+  {\nu/ \sqrt{2} \over  \sin r} (\varphi_{1} -\varphi_{3})=0\,,
\nonumber
\\
- \left ( \cosh t {  \partial \over \partial t}+ 2
 \sinh  t \right )  \varphi_{3}+ \left ( {  \partial \over \partial r}+{1\over \tan r}\right )
 \varphi_{3}+  {\nu/\sqrt{2} \over  \sin r} \varphi_{2} =0\,.
\label{A.20}
\end{eqnarray}

Separating in all functions a special  factor
\begin{eqnarray}
\varphi_{j} = {1 \over \cosh^{2} t} \; {1 \over \sin r } \; F_{j}
\; , \nonumber \label{A.21}
\end{eqnarray}

\noindent we get more simple  equations (for convenience on the
left  some extra numeration is added)
\begin{eqnarray}
 (1')  \left ( {  \partial \over \partial r}+{1\over \tan r} \right ) F_{2}+
 {\nu / \sqrt{2}  \over \sin r} (F_{1} +F_{3})=0,
\nonumber
\\
(2')  -  \cosh t  {  \partial \over \partial t} \,F_{1}
 -  {  \partial \over \partial r}  F_{1}-
 {\nu /\sqrt{2} \over  \sin r} \,F_{2} =0\,,
\nonumber \end{eqnarray}
\begin{eqnarray}
 (3')  -  \cosh t {
\partial \over
\partial t} \, F_{2}+ {\nu/ \sqrt{2} \over  \sin r} (F_{1}
-F_{3})=0\,, \nonumber
\\
(4')  - \cosh t {  \partial \over \partial t}\, F_{3}+
 {  \partial \over \partial r}
 F_{3}+  {\nu/\sqrt{2} \over  \sin r} F_{2} =0.
\label{A.22}
\end{eqnarray}

Let us sum and subtract equations  $(2')$ and $(4') $in (\ref{A.22}) and allow
equation  $(3')$ -- this results
(introducing intermediate parameter
 $\nu/ \sqrt{2} = b$)
\begin{eqnarray}
(2')+(4')\;\;\; -  \cosh t {  \partial \over \partial t}
\left(F_{1}
  +  F_{3}\right)
   -  {  \partial \over \partial r}  \left(  F_{1}-
  F_{3}\right) =0\,,\label{A.23a}
  \\
  -(2')+('4)\qquad   \cosh t {  \partial \over \partial t}  \left(
F_{1}-  F_{3}\right)
 + {  \partial \over \partial r}   \left( F_{1}+
  F_{3}\right)+
 {2b \over  \sin r} \,F_{2} =0\,.
 \label{A.23b}
 \end{eqnarray}

Let us prove that eq.  $(1')$ in  (\ref{A.22}) can be derived from three remaining equations.
To this  end,  we should differentiate in time equation   $(1')$ in (\ref{A.22}):
\begin{eqnarray}
\left ( {  \partial \over \partial r}+{1\over \tan r} \right
){\partial \over \partial t}\, F_{2}+
 {b \over \sin r} {\partial \over \partial t} \left(F_{1} +F_{3}\right)=0\, ;
\nonumber
\end{eqnarray}
and then let us take into account  equation  $(3')$ from  (\ref{A.22}), this gives
\begin{eqnarray}
\left ( {  \partial \over \partial r}+{1\over \tan r} \right ) {b
\over  \cosh t\sin r} (F_{1} -F_{3})
+
 {b \over \sin r} {\partial \over \partial t} \left(F_{1} +F_{3}\right)=0\, ,
\nonumber
\end{eqnarray}

\noindent which is equivalent to
\begin{eqnarray}
{b \over  \cosh t\sin r}   {  \partial \over \partial r} (F_{1}
-F_{3}) +  {b \over \sin r} {\partial \over \partial t}
\left(F_{1} +F_{3}\right)=0 .
\nonumber
\end{eqnarray}

\noindent
It remains to allow for eq. (\ref{A.23a})
\begin{eqnarray}
{  \partial \over \partial t}   (F_{1}
  +  F_{3})= -  {1\over \cosh t} {  \partial \over \partial r} (   F_{1}-
  F_{3}) \,,
\nonumber
\end{eqnarray}
so we  arrive at identity
$ 0 \equiv 0$.
Further, we may use only three independent equations
\begin{eqnarray}
 {  \partial \over \partial t} \, F_{2}=  {b \over
\cosh t\sin r} (F_{1} -F_{3})\,,
  \cosh t {  \partial \over \partial t}
\left(F_{1}
  +  F_{3}\right) +  {  \partial \over \partial r}  \left(  F_{1}-
  F_{3}\right) =0\,,
\nonumber
\\
    \cosh t {  \partial \over \partial t}  \left(
F_{1}-  F_{3}\right)
 + {  \partial \over \partial r}   \left( F_{1}+
  F_{3}\right)+
 {2b \over  \sin r} \,F_{2} =0\,.
 \label{A.24c}
 \end{eqnarray}

One can exclude  $F_{2}$ from the third equation in (\ref{A.24c}). To this end, it suffices to differentiate
(\ref{A.24c}) in time and then to take into account expression for
$\partial _{t}F_{2}$  from the first equation in  (\ref{A.24c}):
\begin{eqnarray}
 \cosh t {  \partial \over \partial t}\cosh t  {  \partial \over \partial t} (F_{1}-   F_{3})
 +   \cosh t {  \partial \over \partial t} {  \partial \over \partial r} (  F_{1}+
  F_{3} )+
 {2b^{2} \over \sin ^{2}r} (F_{1} -F_{3}) =0\,.
\nonumber
\end{eqnarray}

\noindent
So, instead  (\ref{A.24c})  one can use other (equivalent) system
$$
  {  \partial \over \partial t} \, F_{2}=  {b \over  \cosh t\sin r} (F_{1} -F_{3})\,,
$$
$$
  \cosh t {  \partial \over \partial t} \left(F_{1}
  +  F_{3}\right) = -  {  \partial \over \partial r}  (  F_{1}-   F_{3} )  \,,
$$
\begin{eqnarray}
 \cosh t {  \partial \over \partial t}\cosh t  {  \partial \over \partial t} (F_{1}-   F_{3})
 +   \cosh t {  \partial \over \partial t} {  \partial \over \partial r} (  F_{1}+
  F_{3} )+
 {2b^{2} \over \sin ^{2}r} (F_{1} -F_{3}) =0 \,.
\label{A.25c}
\end{eqnarray}

\noindent
With the use of the second equation in (\ref{A.25c}), one  excludes the combination
$(F_{1}+F_{3})$ from (\ref{A.25c}):
\begin{eqnarray}
  {  \partial \over \partial t} \, F_{2}=  {b \over  \cosh t\sin r} (F_{1} -F_{3})\,,
\nonumber\\
  \cosh t {  \partial \over \partial t} \left(F_{1}
  +  F_{3}\right) = -  {  \partial \over \partial r}  (  F_{1}-   F_{3} )  \,,
\nonumber
\\
 \left (  \cosh t {  \partial \over \partial t}\cosh t  {  \partial \over \partial t}
 +    {  \partial \over \partial r}
 {  \partial \over \partial r}   +
 {2b^{2} \over \sin ^{2}r}  \right ) (F_{1} -F_{3}) =0 \,.
\label{A.26c}
\end{eqnarray}

\noindent
Let us simplify notation be introducing other functions:
\begin{eqnarray}
F = F_{1} + F_{3}  \; , \quad G = F_{1}- F_{3}
 \; ,
\nonumber
\end{eqnarray}

\noindent then  (\ref{A.26c}) reads
\begin{eqnarray}
  \cosh t {  \partial \over \partial t} \, F_{2}=  {b \over  \sin r}\, G\,,
 \quad
  \cosh t {  \partial \over \partial t} \,F
= - {  \partial \over \partial r} \, G  \,,
\nonumber
\\
 \left (  -\cosh t  \,{  \partial \over \partial t}\cosh t  \,{  \partial \over \partial t}
  + {  \partial^{2}  \over \partial r^{2}} -
{2b^{2} \over  \sin^{2} r} \right )  G =0\,.\label{A.28}
\end{eqnarray}

It is convenient to introduce a new coordinate instead of $t$ ($\sinh t = x$ is an intermediate variable):
\begin{eqnarray}
\cosh t {  d  \over d  t} = {d \over d \tau },
\quad
 d \tau = {dt
\over \cosh t } = {d x \over 1 + x^{2}},
\quad
 \tau =   \arctan (\sinh t) \; .
 \label{tau}
 \end{eqnarray}

\noindent
Therefore, the system (\ref{A.26c}) can be presented in the form
\begin{eqnarray}
 \left (  -{  \partial ^{2}\over \partial \tau^{2} }
 + {  \partial^{2}  \over \partial r^{2}} -
{2b^{2} \over  \sin^{2} r} \right )  G =0\,,
\nonumber
\\
   {  \partial \over \partial \tau} \, F_{2}=  {b \over  \sin r}\, G\,, \quad
   {  \partial \over \partial \tau} \,F
= - {  \partial \over \partial r} \, G  \, . \label{A.30b}
\end{eqnarray}

Equation for  $G(t,r)$  is solved by separation the variables:
\begin{eqnarray}
G= T(\tau) R(r)\,,
\quad
  {1 \over T(\tau)}  {  \partial ^{2}\over \partial \tau^{2} } T(\tau)
 =  - \omega^{2}\,,
 \nonumber\\
 {1 \over R(r)}\left ( {  \partial^{2}  \over \partial r^{2}} -
{j(j+1) \over  \sin^{2} r} \right )  R(r) =  - \omega^{2};
\nonumber
\end{eqnarray}

\noindent from this it follows
\begin{eqnarray}
T( \tau ) = e^{-i\omega \tau } ,
\quad
 \left ( {  \partial^{2}
\over \partial r^{2}}  + \omega^{2} - {j(j+1) \over  \sin^{2} r}
\right ) R(r) =0\,. \label{A.31}
\end{eqnarray}

To treat the equation for $R(r)$, let us introduce the new variable
$
z = 1 - e^{-2ir}$; $z$ runs along the closed circle of unit length
in complex plane. Allowing for relations
\begin{eqnarray}
{d \over d r} = 2i (1 - z) {d \over d z} \, ,  \quad {1 \over
\sin^{2} r } = -{4(1-z) \over z^{2} }   \, ,
\nonumber
\end{eqnarray}

\noindent we derive
\begin{eqnarray}
4(1-z)^{2}{d ^{2} f\over dz^{2}} \,  -4 (1-z){d f\over dz}
-
\omega^{2} f -{4(1-z)\nu^{2} \over z^{2} }\, f = 0 ,
\nonumber
\end{eqnarray}

\noindent and further through the substitution
 $R = z^{a} (1-z)^{b} f (z)$ the problem reduces to
\begin{eqnarray}
z (1-z) {d^{2} f \over dz^{2}} + [ 2a -(2a+2b+1)z]\; {d f\over d
z}
\nonumber\\
 + \left [ {\omega^{2} \over 4}-(a+b)^{2} +(a(a-1)-\nu^{2}){1 \over
z}
+ (b^{2}-{\omega^{2} \over 4}){1 \over 1-z} \right ] f = 0 \; .
\nonumber
\end{eqnarray}
\noindent With requirements
\begin{eqnarray}
a = j+1 , \;- j \, , \quad b = \pm {\omega \over 2}
\nonumber
\end{eqnarray}

\noindent we get equation of hypergeometric type
\begin{eqnarray}
z (1-z) {d^{2} f \over dz^{2}} + \left [ 2a -(2a+2b+1) z\right] \,
{d f \over d z}
 - [(a+b)^{2} - {\omega^{2} \over 4} ] \,f =0
\label{A.34}
\end{eqnarray}

\noindent with parameters
\begin{eqnarray}
\gamma = 2a \, , \quad \alpha = a+b  - {  \omega \over 2} \,,
\quad \beta = a+b  + {\omega \over 2} \, .
\nonumber
\end{eqnarray}

\noindent The function  $R$ is specified by
\begin{eqnarray}
R = z^{a} (1- z)^{b} f(z)
= \left (\,  2i \sin r e^{-i r } \,
\right ) ^{a} \, \left ( \,  1 -  2i \sin r e^{-i r }  \, \right )
^{b} \, f(z)\, ;
\nonumber
\end{eqnarray}

\noindent this solution is finite at the points
 $r =0$ and $r = \pi$ only at positive  $a$:
$
a = j+1$, and at  $b = - \omega /2 $, then hypergeometric series can be restricted to polonimials
(by physical grounds one should assume $\omega > 0$):
\begin{eqnarray}
\alpha = j+1  -  \omega   = - n  = \{ 0, -1, -2, ... \; \} \quad
\Longrightarrow
\quad
\omega = n + 1 + j \; .
\nonumber
\end{eqnarray}

\noindent Thus, the appropriate solutions are given by
\begin{eqnarray}
R = z^{a} (1- z)^{b} f(z)
\nonumber\\
= \left (\,  2i \sin r e^{-i r } \,
\right ) ^{a}  \left ( \,  1 -  2i \sin r e^{-i r }  \, \right )
^{b}  f(z)\, ,
\nonumber
\\
f(z) =  F (-n, \,j +1, \,2j +2; \; 2i \sin r e^{-i r } ) \, ;
\label{A.37a}
\end{eqnarray}

\noindent quantization is determined by the formula
\begin{eqnarray}
\omega = n + 1 + j \, , \quad j = 0, 1, 2, ....,
\quad
 n = 0, 1,
2, ... ; \label{A.37b}
\end{eqnarray}

\noindent or in usual units
($\rho$ ia the curvature radius)
\begin{eqnarray}
\omega = {c \over \rho } \; ( n + 1 + j )\, .
\nonumber
\end{eqnarray}

Let us turn to the first two equations in  (\ref{A.28}) --  they can be presented as
\begin{eqnarray}
   {  \partial \over \partial \tau } \, F_{2}=  {b \over  \sin r}\, e^{-i\omega \tau }R(r) \,,
   \quad
  {  \partial \over \partial \tau } \,F
= - {  \partial \over \partial r} \, e^{-i\omega \tau }R(r)   \,;
\nonumber
\end{eqnarray}

\noindent from whence it follows
\begin{eqnarray}
    F_{2} (t,r) = -  {1 \over i \omega} \; e^{-i\omega \tau } \; {b \over  \sin r}\, R(r) \,,
  \quad
  F(t,r) = + {1 \over i \omega} \; e^{-i\omega \tau } \; {  d  \over d r} R(r)   \,.
\label{A.38}
\end{eqnarray}

\section{Relation between Majorana--Oppenheimer  and Duffin--Kemmer--Petiau formalisms;
  electromagnetic waves of magnetic and electric types}

In contrast to Majorana--Oppenheimer approach, the DKP formalisms
permits us to follow gauge degrees of freedom of the electromagnetic field.
Let us relate these descriptions.
It is convenient to start with  yet known structures for  electromagnetic complex 3-vector
 and 10-dimensional DKP field function
 \begin{eqnarray}
\Psi  = e^{-i \omega t} \left | \begin{array}{l}
0 \\
\varphi_{1} D_{-1 }
\\
\varphi_{2} D_{0 }  \\
\varphi_{3} D_{+1 }
\end{array} \right |,
\nonumber\\
\Phi   = e^{-i\omega t}  \; [ f_{1} D_{0 }  ; f_{2} \; D_{-1} ,
f_{3} \;  D_{0} ,  f_{4} \; D_{+1};
\nonumber
\\
 f_{5} \; D_{-1} , f_{6} \;  D_{0} ,  f_{7} \; D_{+1} ;
 \nonumber
\\
f_{8} \; D_{-1} ,  f_{9} \; D_{0} ,f_{10} \; D_{+1} ]\,.
\label{A.41a}
\end{eqnarray}

\noindent In accordance with definitions, there must exist
relationships
\begin{eqnarray}
E_{1}= F_{01 }   \, ,\quad E_{2} =  F_{02 } \, , \quad E_{3} =
F_{03 }   \, ,
\nonumber
\\
B_{1} = -  F_{23 }  \, , \quad B_{2} = - F_{31 }  \, , \quad
B_{3} = - F_{12 } \;;
\label{A.41b}
\end{eqnarray}

\noindent   to avoid misunderstanding we denote $(t,r)$-constituents in Majorana--Oppenheimer
picture by $\varphi_{j}(t,r)$:
\begin{eqnarray}
 \varphi_{1}    D_{-1} = E_{1} + i B_{1} =
f_{5}  D_{-1} -i \; f_{8} \; D_{-1}   \; ,
\nonumber
\\
\varphi_{2}  D_{0}  = E_{2} + i B_{2} =
  f_{6}  \; D_{0} - i f_{9} \;D_{0}   \; ,
\nonumber
\\
 \varphi_{3} \, D_{+1}   =   E_{3} + i B_{3}
 = f_{7} \, D_{+1}  -i f_{10} \,D_{+1}  \, .
\label{A.41c}
\end{eqnarray}

\noindent From  (\ref{A.41c}) it follows
\begin{eqnarray}
 \varphi_{2}  =  f_{6}   - i \, f_{9}   \, , \quad
\varphi_{1}  =  f_{5}   -i \, f_{8}    \, ,
\quad
  \varphi_{3}  =    f_{7}    -i \,  f_{10}  \, .
\label{A.42}
\end{eqnarray}

Allowing for restrictions, referred to spacial parity (see in \cite{Book-2011}):
\begin{eqnarray}
 P = (-1)^{j+1} \, , \;\; f_{6} = 0 \, , \; \; f_{7}
= - f_{5}\,,\;\; f_{10} = + f_{8}  ;
\nonumber\\
P = (-1)^{j} \, , \quad
 f_{9} = 0\, , \;  \; f_{7} = + f_{5}\, , \;\;
f_{10} = - f_{8}\; ,
\nonumber
\end{eqnarray}

\noindent
instead of (\ref{A.42}) we   introduce two classes of solutions:
\begin{eqnarray}
 P = (-1)^{j+1} \, , \qquad
 \varphi_{2}  =    - i \, f_{9}   \, ,
\nonumber\\
 \varphi_{1}  =  f_{5}   -i \, f_{8}    \, , \quad
  \varphi_{3}  =    - f_{5}    -i \,  f_{8}  \, ;
\label{A.43a}
\\
P = (-1)^{j} \, , \qquad
 \varphi_{2}  =  f_{6}      \, ,
 \nonumber\\
\varphi_{1}  =  f_{5}   -i \, f_{8}    \, , \quad
  \varphi_{3}  =    f_{5}    +i \,  f_{8}  \, ;
\label{A.43b}
\end{eqnarray}

\noindent inverse relations are
\begin{eqnarray}
 P = (-1)^{j+1} \, , \qquad
 f_{9} = i  \varphi_{2}         \, ,
 \nonumber\\
 f_{5} = {1 \over 2}  (\varphi_{1}- \varphi_{3} ) \,, \quad
 f_{8} = {i \over 2}  (\varphi_{1}+ \varphi_{3} )
\,; \label{A.44a}
\\
P = (-1)^{j} \, , \qquad
   f_{6}  = \varphi_{2}       \, ,
   \nonumber\\
    f_{5} = {1  \over 2}  (\varphi_{1}+ \varphi_{3} )\,,
\quad
   f_{8} =  {  i \over 2}  (\varphi_{1}- \varphi_{3} ) \, .
\label{A.44b}
\end{eqnarray}

It should be stressed that solutions of Majorana--Oppenheimer equation
which  differ  in the imaginary unit $i$,   from physical standpoint represent
substantially different electromagnetic fields (in fact, one should take in mind
complex conjugated equation and its solutions).

After translating description to other functions
\begin{eqnarray}
\varphi_{j} = {1 \over \cosh^{2}t \sin r} F_{j}\,,
\nonumber\\
 F = F_{1}
+ F_{3}  \, , \quad G = F_{1}- F_{3}
 \, ,
\nonumber
\end{eqnarray}

\noindent
the formulas  (\ref{A.44a})  and (\ref{A.44b}) will read
\begin{eqnarray}
 P = (-1)^{j+1} \, , \qquad
 f_{9} = i  \varphi_{2}         \, ,
 \nonumber\\
 f_{5} = {1 \over 2}  {1 \over \cosh^{2}t \sin r} G \,,
 \nonumber\\
 f_{8} = {i \over 2}  {1 \over \cosh^{2}t \sin r} F
\, ;
\label{A.45a}
\\[4mm]
P = (-1)^{j} \, , \qquad
   f_{6}  = \varphi_{2}       \, ,
   \nonumber\\
    f_{5} = {1  \over 2}
   {1 \over \cosh^{2}t \sin r} F \,,
\nonumber\\
   f_{8} =  {  i \over 2}  {1 \over \cosh^{2}t \sin r} G  \, .
\label{A.45b}
\end{eqnarray}

Now we should relate the systems of $(t,r)$-equations in these two formalism
(we omit here all details of deriving such equations in DK formalism -- it is separate and rather
laborious task).

First, let us consider the system of $(t,r)$-equations in DK approach
for states with parity (we  omit all  details
concerning deriving these equations, they are part of another paper
in preparation on treating
the massive spin one particle in expending Universe)

 $ P = (-1)^{j+1},$
\begin{eqnarray}
-\cosh t  {\partial \over \partial t} f_{5}  - 2\sinh t f_{5}
\nonumber\\
+ i(
{\partial \over \partial r}  + { 1 \over  \tan r} ) f_{8}+ { i \nu
/\sqrt{2} \over \sin r }  f_{9}=0 \; ,
\nonumber
\\
\cosh t ( { \partial \over \partial t} f_{2} -  f_{5} ) + \sinh t
f_{2} = 0 \; ,
\nonumber
\\
-  \cosh t f_{8} - i ( {\partial \over \partial r}   + { 1 \over
\tan r } )  f_{2}  =0 \; ,
\nonumber\\
 - \cosh t  f_{9}  + { i \sqrt{2}
\nu   \over \sin r }   f_{2}=0 \; . \label{A.46}
\end{eqnarray}

\noindent Let us translate (\ref{A.46}) to notation according to  (\ref{A.44a}),
 which results in
\begin{eqnarray}
{1\over 2} \left (  \cosh t  {\partial \over \partial t}  + 2\sinh
t \right )   (\varphi_{1}- \varphi_{3} )
\nonumber\\
+ (   {\partial \over
\partial r}  + { 1 \over  \tan r} ) {1 \over 2}  (\varphi_{1}+
\varphi_{3} ) + {  \nu /\sqrt{2} \over \sin r }    \varphi_{2} =0
 ,
\nonumber
\\(\cosh t  { \partial \over \partial t}  + \sinh t   ) f_{2}
 - \cosh t  {1 \over 2}  (\varphi_{1}- \varphi_{3} )
= 0 \, ,
\nonumber
\end{eqnarray}
\begin{eqnarray}
  \cosh t {1 \over 2}  (\varphi_{1}+ \varphi_{3} )  +  ( {\partial \over \partial r}   + { 1 \over
\tan r } )  f_{2}  =0 \; ,
\nonumber\\[3mm]
 -\cosh t  \;   \varphi_{2}   + {\sqrt{2}  \nu   \over \sin r }   f_{2}=0 \; ;
\label{A.47}
\end{eqnarray}

\noindent The last equation permits us to exclude  $f_{2}$ (it is referred to 4-potential
of electromagnetic
field)
\begin{eqnarray}
f_{2} =  { \cosh t \; \sin r \over \sqrt{2} \nu } \varphi_{2}
\nonumber
\end{eqnarray}

\noindent from second and third equations in (\ref{A.47}):
\begin{eqnarray}
 \left (  \cosh t  {\partial \over \partial t}  + 2\sinh t
\right )   (\varphi_{1}- \varphi_{3} )
\nonumber\\
 + (   {\partial \over
\partial r}  + { 1 \over  \tan r} )
  (\varphi_{1}+ \varphi_{3} ) + {  \sqrt{2} \nu \over \sin r }    \varphi_{2} =0 \, ,
\nonumber
\\[3mm]
-( \cosh t  { \partial \over \partial t}  + 2 \sinh t )
\varphi_{2}
+
  { \nu /\sqrt{2} \over \sin r}   (\varphi_{1}- \varphi_{3} )
= 0 \, ,
\nonumber\\[3mm]
    {\nu / \sqrt{2}  \over \sin r}  (\varphi_{1}+ \varphi_{3} )  + ( {\partial \over \partial r}   + { 2 \over
\tan r } )      \varphi_{2}
  =0 \, .
\label{A.49}
\end{eqnarray}

It should be noted that acting on  the last equation in (\ref{A.49}) by the operator
 $
 ( \cosh t \partial _{ t} + 2 \sinh t  )
$ :
\begin{eqnarray}
    {\nu / \sqrt{2}  \over \sin r}
    \left ( \cosh t {\partial \over \partial t} + 2 \sinh t \right )
     (\varphi_{1}+ \varphi_{3} )
\nonumber\\
     + ( {\partial \over \partial r}   + { 2 \over
\tan r } )    \left ( \cosh t {\partial \over \partial t} + 2
\sinh t \right )   \varphi_{2}
  =0 \, ,
\nonumber
\end{eqnarray}

\noindent and then allowing for second equation in (\ref{A.49}), we arrive at the equation
\begin{eqnarray}
    \left ( \cosh t {\partial \over \partial t} + 2 \sinh t \right )
     (\varphi_{1}+ \varphi_{3} )
     \nonumber\\
     + ( {\partial \over \partial r}   + { 1 \over
\tan r } )   (\varphi_{1}- \varphi_{3} )
  =0 \,.
\label{A.50}
\end{eqnarray}

\noindent This means that the system  (\ref{A.49}) is equivalent to the following one
\begin{eqnarray}
 \left (  \cosh t  {\partial \over \partial t}  + 2\sinh t
\right )   (\varphi_{1}- \varphi_{3} )
\nonumber\\
+ (   {\partial \over
\partial r}
+ { 1 \over  \tan r} )
  (\varphi_{1}+ \varphi_{3} ) + {  \sqrt{2} \nu \over \sin r }    \varphi_{2} =0 \, ,
\nonumber
\\[3mm]
-( \cosh t  { \partial \over \partial t}  + 2 \sinh t )
\varphi_{2}
 +
  { \nu /\sqrt{2} \over \sin r}   (\varphi_{1}- \varphi_{3} )
= 0 \, ,
\nonumber
\\[3mm]
    \left ( \cosh t {\partial \over \partial t} + 2 \sinh t \right )
     (\varphi_{1}+ \varphi_{3} )
         + ( {\partial \over \partial r}   + { 1 \over
\tan r } )   (\varphi_{1}- \varphi_{3} )
  =0 \, .
\label{A.51}
\end{eqnarray}

Now we are ready to demonstrate equivalence of equations derived in the frames of DKP
approach (\ref{A.49}), (\ref{A.51}) to equations arising in  Majorana--Oppenheimer  approach (\ref{A.20}).
To this end, let us turn to  the system (\ref{A.20}) with excluded first equation (which is
a consequence of these three):
\begin{eqnarray}
 \left ( \cosh t  {  \partial \over \partial t}+ 2
 \sinh t  \right ) \, \varphi_{1}
 \nonumber\\
 + \left ( {  \partial \over \partial r} +{1\over \tan r} \right ) \varphi_{1} +
 {\nu /\sqrt{2} \over  \sin r} \,\varphi_{2} =0\,,
\nonumber
\\[3mm]
- \left ( \cosh t {  \partial \over \partial t} + 2
 \sinh t \right ) \varphi_{2}+  {\nu/ \sqrt{2} \over  \sin r} (\varphi_{1} -\varphi_{3})=0,
\nonumber
\end{eqnarray}
\begin{eqnarray}
- \left ( \cosh t {  \partial \over \partial t}+ 2
 \sinh  t \right )  \varphi_{3}
  + \left ( {  \partial \over \partial r}+{1\over \tan r}\right )
 \varphi_{3}+  {\nu/\sqrt{2} \over  \sin r} \varphi_{2} =0\,.
\label{A.52}
\end{eqnarray}

Note that  the second equations in (\ref{A.51}) and  (\ref{A.52}) coincide.
Next, let us sum equations one and three in  (\ref{A.52}):
\begin{eqnarray}
 \left ( \cosh t  {  \partial \over \partial t}+ 2
 \sinh t  \right ) \, (\varphi_{1} - \varphi_{3})
 \nonumber\\
 + \left ( {  \partial \over \partial r} +{1\over \tan r} \right ) (\varphi_{1}+\varphi_{3})  +
 {\sqrt{2} \nu  \over  \sin r} \,\varphi_{2} =0\,,
\nonumber
\end{eqnarray}

\noindent this one  coincides with the first equation in  (\ref{A.51}).

Now, let us subtract the first equation  from the third equation in  (\ref{A.52}):
\begin{eqnarray}
 \left ( \cosh t  {  \partial \over \partial t}+ 2
 \sinh t  \right ) \, (\varphi_{1} + \varphi_{3})
  + \left ( {  \partial \over \partial r} +{1\over \tan r} \right ) (\varphi_{1} -\varphi_{3})
  =0\, ,
\nonumber
\end{eqnarray}

\noindent this one coincides with the third equation in (\ref{A.51}).

Thus, for states of electromagnetic field with parity
 $(-1)^{j+1}$, DK- and MO-approaches give equivalent systems in
 $(t,r)$-variables.

Now let us turn to the system of $(t,r)$-equations derived in DK formalism for states with
the opposite  parity

\vspace{2mm} $ P=(-1)^{j} $
\begin{eqnarray}
(  {\partial \over \partial r}    + {2 \over  \tan r}   )  f_{6}
+{ 2\nu  \over \sin r } \;
 f_{5}   =0 \, ,
\nonumber
\\
-\cosh t  ({\partial \over \partial t} + 2\tanh  t  ) f_{5}
+
 i(   {\partial \over \partial r}  + { 1 \over  \tan r} )  f_{8}    =0 \, ,
\nonumber
\\
-\cosh t (  {\partial \over \partial t}    +  2\tanh  t  ) f_{6}-
{ 2i  \nu  \over \sin r }  f_{8}    =0 \, ; \label{A.53a}
\end{eqnarray}
\begin{eqnarray}
\cosh t  ( { \partial \over \partial t} + \tanh t )  f_{2}
+
 { \nu  \over \sin r }  f_{1}  -   \cosh t f_{5} = 0 \, ,
\nonumber
\\
- \cosh t ( {\partial \over \partial t}  + \tanh  t)  f_{3}
 +
 {\partial \over \partial r} f_{1}   +   \cosh t f_{6} =0\, ,
\nonumber
\\
 i ({\partial \over \partial r}    + { 1 \over  \tan r} )  f_{2} +
{ i  \nu  \over \sin r }  f_{3} +  \cosh t  f_{8} =0 \, ,
\label{A.53b}
\end{eqnarray}

\noindent where  $\nu =\sqrt{j(j+1)}/\sqrt{2}$).

First let us consider the first three equations  (\ref{A.53a}),
in which only related to tensors functions enter. Translate them to
new function according to (\ref{A.44b}):
\begin{eqnarray}
(  {\partial \over \partial r}    + {2 \over  \tan r}   )
\varphi_{2} +{ \nu  \over \sin r } \;
   (\varphi_{1}+ \varphi_{3} )   =0 \; ,
\nonumber
\\
-\cosh t  ({\partial \over \partial t} + 2\tanh  t  )
(\varphi_{1}+ \varphi_{3} )
-
 (   {\partial \over \partial r}  + { 1 \over  \tan r} )
    (\varphi_{1}- \varphi_{3} )    =0 \; ,
\nonumber
\\
-\cosh t (  {\partial \over \partial t}    +  2\tanh  t  )
\varphi_{2}
 + {   \nu  \over \sin r }    (\varphi_{1}-
\varphi_{3} )    =0 \, . \nonumber
\end{eqnarray}

\noindent
Next, let us translate them to $F_{j}$:
\begin{eqnarray}
\varphi_{j} = {1 \over \cosh^{2}t \sin r} F_{j},
\quad
 F = F_{1}
+ F_{3}  \, , \quad G = F_{1}- F_{3}
 \, ;
\nonumber
\end{eqnarray}

\noindent so we get
\begin{eqnarray}
(  {\partial \over \partial r}    + {1 \over  \tan r}   )  F_{2}
+{ \nu  \over \sin r } \;
   F   =0 \; ,
\nonumber
\\
-\cosh t   {\partial \over \partial t}    F_{2}   + { \nu  \over
\sin r }    G  =0 \; ,
\nonumber
\\
\cosh t  {\partial \over \partial t}  F  +
    {\partial \over \partial r}  G
        =0 \; .
\label{A.56}
\end{eqnarray}

\noindent Note that third equation is  a consequence of the other.
Indeed, differentiating the first one in time
\begin{eqnarray}
(  {\partial \over \partial r}    + {1 \over  \tan r}   )
{\partial \over \partial t} F_{2} +{ \nu  \over \sin r } \,
   {\partial \over \partial t} F   =0 \, ,
\nonumber
\end{eqnarray}

\noindent and allowing for the second, we derive the third
\begin{eqnarray}
(  {\partial \over \partial r}    + {1 \over  \tan r}   )  { \nu
\over \cosh t \sin r }    G +{ \nu  \over \sin r } \,
   {\partial \over \partial t} F   =0 \; \Longrightarrow \;\;
{1 \over \cosh t} {\partial \over \partial r } G + {\partial \over
\partial t } F =0 \,.
\nonumber
\end{eqnarray}

Now, let us turn to the equations  (\ref{A.22}) derived in MO approach ($\nu
=\sqrt{j(j+1)/2}$)
\begin{eqnarray}
  \left ( {  \partial \over \partial r}+{1\over \tan r} \right ) F_{2}+
 {\nu   \over \sin r} (F_{1} +F_{3})=0,
\nonumber
\\
( -  \cosh t {  \partial \over \partial t} \, F_{2}+
{\nu \over  \sin r} (F_{1} -F_{3})=0\,,
\nonumber
\\
 -  \cosh t {  \partial \over \partial t}
(F_{1}
  +  F_{3})
    -  {  \partial \over \partial r}  \left(  F_{1}-
  F_{3}\right) =0\,,
\nonumber
\end{eqnarray}
\begin{eqnarray}
 \cosh t {  \partial \over \partial t}  \left(
F_{1}-  F_{3}\right)
 \nonumber\\
 + {  \partial \over \partial r}   \left( F_{1}+
  F_{3}\right)+
 {2\nu \over  \sin r} \,F_{2} =0\,;
 \label{A.57}
 \end{eqnarray}

\noindent in the functions  $F_{2},\, F, \,G$ it reads
\begin{eqnarray}
 \left ( {  \partial \over \partial r}+{1\over \tan r} \right ) F_{2}+
 {\nu   \over \sin r} F=0\,,
\nonumber
\\
 -  \cosh t {  \partial \over \partial t} \, F_{2}+
{\nu \over  \sin r} G=0\,,
\nonumber
\\
   \cosh t {  \partial \over \partial t} F +
 {  \partial \over \partial r}  G =0\,,
\nonumber
\\
 \cosh t {  \partial \over \partial t}  G
 + {  \partial \over \partial r}   F +
 {2\nu \over  \sin r} \,F_{2} =0\,.
 \label{A.58}
 \end{eqnarray}

\noindent
Note coincidence of the first three equations in (\ref{A.58}) and
 (\ref{A.56})
\begin{eqnarray}
(  {\partial \over \partial r}    + {1 \over  \tan r}   )  F_{2}
+{ \nu  \over \sin r } \;
   F   =0 \; ,
   \nonumber\\
-\cosh t   {\partial \over \partial t}    F_{2}   + { \nu  \over
\sin r }    G  =0\; ,
\nonumber\\
\cosh t  {\partial \over \partial t}  F  +
    {\partial \over \partial r}  G
        =0 \; .
\nonumber
\end{eqnarray}

\noindent
One can easily exclude from the fourth equation in MO system
(\ref{A.58}) the functions  $F$ and $F_{2}$. To this end, it suffices
to differentiate it in time
\begin{eqnarray}
{  \partial \over \partial t}  \cosh t {  \partial \over \partial
t}  G
 + {  \partial \over \partial r} {  \partial \over \partial t}   F +
 {2\nu \over  \sin r} \,{  \partial \over \partial t}  F_{2} =0\,,
\nonumber
\end{eqnarray}

\noindent and then take into account 2-nd and 3-rd equations
in  (\ref{A.58}):
$$
\left ( {  \partial \over \partial t}  \cosh t {  \partial \over
\partial t}
 -{1 \over \cosh t} {\partial ^{2} \over \partial r^{2} }
+ {2 \nu^{2} \over \sin^{2} r}  {1 \over \cosh t} \right ) G =0
\;.
$$

\noindent
Thus, we derive a yet familiar equation for $G(t,r)$:
\begin{eqnarray}
 \cosh t {  \partial \over \partial t}  \cosh t {  \partial \over \partial t} G =
 \left (  {\partial ^{2} \over \partial r^{2} }
- {j(j+1) \over \sin^{2} r}  \right ) G  ,
\label{A.59b}
\end{eqnarray}

\noindent
or
\begin{eqnarray}
\cosh t {  d \over d t } = {d \over d \tau }\,,
\quad
{d^{2} \over d \tau^{2} } G =  \left (  {\partial ^{2} \over
\partial r^{2} } - {j(j+1) \over \sin^{2} r}  \right ) G  \,.
\label{A.59c}
\end{eqnarray}
\noindent
Its solution was found after (\ref{A.31})
\begin{eqnarray}
G = e^{-i\omega \tau } R(r) \; .
\nonumber
\label{A.59d}
\end{eqnarray}

In turn, equation from  (\ref{A.58})  provides us with expressions
for two remaining functions
\begin{eqnarray}
   {\partial \over \partial \tau }    F_{2}  = { \nu  \over \sin r }    G   \, ,\quad
 {\partial \over \partial \tau}  F  = -     {\partial \over \partial r}  G
         \, ;
\nonumber
\end{eqnarray}

\noindent
from these it follows
\begin{eqnarray}
F_{2} =  -{1 \over i\omega}  e^{-i\omega \tau } { \nu  \over \sin
r } R(r) \, , \quad
 F = {1 \over i\omega}  e^{-i\omega \tau } {d
\over dr } R (r) \, . \label{A.59e}
\end{eqnarray}

Now let us  consider three remaining equations from DKP system (\ref{A.53b})
\begin{eqnarray}
(\cosh t   { \partial \over \partial t} + \sinh t )  f_{2}
+
 { \nu  \over \sin r }  f_{1}  -   \cosh t f_{5} = 0 \, ,
\nonumber
\\[3mm]
- (\cosh t  {\partial \over \partial t}  + \sinh  t)  f_{3}
+
 {\partial \over \partial r} f_{1}   +   \cosh t f_{6} =0\, ,
\nonumber
\\[3mm]
  ({\partial \over \partial r}    + { 1 \over  \tan r} )  f_{2} +
{   \nu  \over \sin r }  f_{3} -i   \cosh t  f_{8} =0 \, .
\label{A.60}
\end{eqnarray}

\noindent
One can exclude functions $f_{2},\, f_{3}$ from third equation in  (\ref{A.60}); it suffices to act on it by operator
\begin{eqnarray}
(\cosh t   { \partial \over \partial t} + \sinh t ) \,,
\nonumber
\end{eqnarray}

\noindent and then take into account second equation from (\ref{A.60}), which results in
\begin{eqnarray}
  ({\partial \over \partial r}    + { 1 \over  \tan r} )
  \left (  - { \nu  \over \sin r }  f_{1}  +   \cosh t f_{5}\right  )
   \nonumber\\
    +
    {   \nu  \over \sin r }
\left (   {\partial \over \partial r} f_{1}   +  \cosh t f_{6}
\right )
\nonumber\\
- i   (\cosh t   { \partial \over \partial t} + \sinh t )  \cosh t
f_{8} =0 \, .
\nonumber
\end{eqnarray}

\noindent
In more detailed form it reads
\begin{eqnarray}
-{\nu \over \sin r} {\partial \over \partial r} f_{1} + {\nu \cos
r \over \sin^{2} r} f_{1} -  {\nu \cos r \over \sin^{2} r} f_{1}
\nonumber\\
+ \cosh t ( {\partial \over \partial r} +{1 \over \tan r}) f_{5}
\nonumber\\
+ {\nu \over \sin r} {\partial \over \partial r} f_{1}
+ \cosh t
{\nu \over \sin r} f_{6}
\nonumber\\
 -
 i   (\cosh t   { \partial \over \partial t} + \sinh t )  \cosh t
f_{8} =0 \, .
\nonumber
\end{eqnarray}

\noindent Terms with $f_{1} $ cancel out each other and we get
\begin{eqnarray}
\cosh t ( {\partial \over \partial r} +{1 \over \tan r}) f_{5}
 + \cosh t {\nu \over \sin r} f_{6}
 \nonumber\\
 -
 i   (\cosh t   { \partial \over \partial t} + \sinh t )  \cosh t  f_{8} =0 \; .
\label{A.61b}
\end{eqnarray}

Now let translate this equation to other notation
\begin{eqnarray}
\underline{P = (-1)^{j} }\, , \quad
   f_{6}  = \varphi_{2}       \, ,
   \nonumber\\
    f_{5} = {1  \over 2}  (\varphi_{1}+ \varphi_{3} )\,,
\quad
   f_{8} =  {  i \over 2}  (\varphi_{1}- \varphi_{3} ) \, ;
\nonumber
\end{eqnarray}
so we get
\begin{eqnarray}
\cosh t ( {\partial \over \partial r} +{1 \over \tan r})
(\varphi_{1}+ \varphi_{3} )
 + \cosh t {2\nu \over \sin r} \varphi_{2}
 \nonumber\\
 +
    (\cosh t   { \partial \over \partial t} + \sinh t )  \cosh t
    (\varphi_{1} - \varphi_{3})  =0 \, .
\label{A.62a}
\end{eqnarray}

\noindent Next, we make substitutions
\begin{eqnarray}
\varphi_{2} = { F_{2} \over \cosh^{2} t \sin r } \,, \quad
(\varphi_{1}+ \varphi_{3} ) = {  F \over \cosh^{2} t \sin r} ,
\nonumber\\
 (\varphi_{1}- \varphi_{3} ) = {  G \over \cosh^{2} t \sin
r} \, ;
\nonumber
\end{eqnarray}

\noindent which results in
\begin{eqnarray}
{1 \over \cosh t \sin r}   {\partial \over \partial r}  F
 + {1 \over \cosh t}  {2\nu \over \sin^{2}  r} F_{2}
 +{1 \over \sin r}
    (\cosh t   { \partial \over \partial t} + \sinh t )  {1 \over \cosh t} G =0 \, ,
\nonumber
\end{eqnarray}

\noindent which leads to
\begin{eqnarray}
   {\partial \over \partial r}  F
 +   {2\nu \over \sin  r} F_{2} +
    \cosh t  { \partial \over \partial t} G  =0 \,.
\label{A.62c}
\end{eqnarray}

\noindent
In this equation, one can exclude functions $F$ and $
F_{2}$.  To this end, it suffices to differentiate it in time
\begin{eqnarray}
   {\partial \over \partial r}  \cosh t   {\partial  \over \partial t}  F
 +   {2\nu \over \sin  r}  \cosh t   {\partial  \over \partial t}  F_{2}
  +
  \cosh t   {\partial  \over \partial t}  \cosh t  { \partial \over \partial t} G  =0
\nonumber
\end{eqnarray}

\noindent and allow for two equations from  (\ref{A.58}):
\begin{eqnarray}
\cosh t   {\partial \over \partial t}    F_{2}  = { \nu  \over
\sin r }    G \, , \quad \cosh t  {\partial \over \partial t}  F
= -    {\partial \over \partial r}  G         \, .
\nonumber
\end{eqnarray}

\noindent In result, we arrive at yet known equation (\ref{A.59b})
\begin{eqnarray}
\left (  -  {\partial ^{2} \over \partial r^{2} }
 +   {2\nu ^{2}\over \sin ^{2}  r}   +
  \cosh t   {\partial  \over \partial t}  \cosh t  { \partial \over \partial t} \right ) G  =0 \, .
\nonumber
\end{eqnarray}

Let us turn back to the first two equations in  (\ref{A.60}):
\begin{eqnarray}
(\cosh t   { \partial \over \partial t} + \sinh t )  f_{2} +
 { \nu  \over \sin r }  f_{1}  -   \cosh t f_{5} = 0 \, ,
\nonumber\\
- (\cosh t  {\partial \over \partial t}  + \sinh  t)  f_{3}   +
 {\partial \over \partial r} f_{1}   +   \cosh t f_{6} =0\, ,
\nonumber
\end{eqnarray}

\noindent
translating them  to other functions
\begin{eqnarray}
   f_{6}  = \varphi_{2}  = {F_{2} \over \cosh^{2} t \sin r }       \, ,
   \nonumber\\
    f_{5} = {1  \over 2}  (\varphi_{1}+ \varphi_{3} ) =
    {1 \over 2}    {F \over \cosh^{2} t \sin r   } \,,
\nonumber
\end{eqnarray}

\noindent we get
\begin{eqnarray}
(\cosh t   { \partial \over \partial t} + \sinh t )  f_{2}
+
 { \nu  \over \sin r }  f_{1}  -    {1 \over 2}    {F \over \cosh  t \sin r   } = 0 \, ,
\nonumber
\\[3mm]
 (\cosh t  {\partial \over \partial t}  + \sinh  t)  f_{3}
  -
 {\partial \over \partial r} f_{1}   -    {F_{2} \over \cosh  t \sin r }   =0\,.
\label{A.63b}
\end{eqnarray}

\noindent
After separating the factor
\begin{eqnarray}
f_{1} = {g_{1}  \over \cosh t  } \, ,  \;\; f_{2} = {g_{2}
\over \cosh t  } \, ,
\;\;
 f_{3} = {g_{3}  \over \cosh t  } \, ,
\label{A.63c}
\end{eqnarray}

\noindent
eqs. (\ref{A.63b}) read
\begin{eqnarray}
\cosh t {\partial \over \partial t} g_{2} + {\nu \over \sin r}
g_{1} ={ F / 2 \over \sin r}  \, ,
\nonumber\\
\cosh t {\partial \over \partial t} g_{3} -  {\partial \over
\partial r} g_{1} = {F _{2}\over \sin r}   \, . \label{A.63d}
\end{eqnarray}

\noindent
In the variable  $\tau$ eqs.  (\ref{A.63d}) become simpler
\begin{eqnarray}
 {\partial \over \partial \tau } g_{2} + {\nu \over \sin r}  g_{1} ={ F / 2 \over \sin r}  \, ,
 \nonumber\\
 {\partial \over \partial \tau } g_{3} -{\partial \over \partial r} g_{1} = {F _{2} \over \sin r}  \, .
\label{A.64}
\end{eqnarray}

However, these equations do not permit us  to calculate the functions  $g_{1}, \,g_{2},\,
g_{3}$ by known  $F_{2}, \,F$; this is what we should expect because of existence of the gauge freedom in electrodynamics.

For instance, one can set  $g_{1}(t,r) =0$, then eqs. (\ref{A.64}) will  read
\begin{eqnarray}
 {\partial \over \partial \tau } g_{2}  ={ F / 2 \over \sin r}  \, , \quad
 {\partial \over \partial \tau } g_{3}  = {F _{2}\over \sin r}  \, ,
\label{A.65}
\end{eqnarray}

\noindent which leads to explicit functions  $g_{2},\, g_{3}$ without any gauge freedom.
In this case, eqs.   (\ref{A.65}) determine, in fact,  wave of electric type in Landau gauge (when
the component
 $
 \Psi_{0}$ of the 4-potential vanishes).

When considering (\ref{A.64}) and  (\ref{A.65})  we should remember on the known expressions for
$F_{2},\, F$ (\ref{A.59e}):
\begin{eqnarray}
F_{2} =  -{1 \over i\omega}  e^{-i\omega \tau } { \nu  \over \sin
r } R(r)  , \;\;  F = {1 \over i\omega}  e^{-i\omega \tau } {d
\over dr } R (r)  .
\nonumber
\end{eqnarray}

To obtain description of the waves of electric type in Lorentz gauge eqs.  (\ref{A.63d}) consistent with
the Lorentz condition. The Lorentz condition looks as follows (see the proof of this formula in  the next section)
\begin{eqnarray}
 - { 2\nu   \over \sin r}      g_{2}     +
  (\cosh t     \partial_{t}  + 2   \sinh t ) g_{1}
    - ( \partial _{r}
  + {2 \over \tan r} ) g_{3}
=0    \, . \label{A.66}
\end{eqnarray}

Excluding here the functions $g_{2}, \,g_{3}$ with the help of (\ref{A.63d})
\begin{eqnarray}
\cosh t {\partial \over \partial t} g_{2} =- {\nu \over \sin r}g_{1} +{ F / 2 \over \sin r}  \, ,
\quad
\cosh t {\partial \over \partial t} g_{3} =
  {\partial \over\partial r} g_{1}+ {F _{2}\over \sin r}  \, ;
\nonumber
\end{eqnarray}

\noindent
we get
\begin{eqnarray}
 - { 2\nu   \over \sin r}  \left (- {\nu \over \sin r}g_{1} +{ F / 2 \over \sin r}  \right )
 \nonumber\\
  +
 \cosh t {\partial \over \partial t} (\cosh t     \partial_{t}  + 2   \sinh t ) g_{1}
 \nonumber\\
 - ( \partial _{r}
  + {2 \over \tan r} )\left (  {\partial \over\partial r} g_{1}+{ F _{2}\over \sin r}  \right )
=0    \, ,
\nonumber
\end{eqnarray}

\noindent
 after transformation it looks
\begin{eqnarray}
 \left [   \cosh t {\partial \over \partial t} (\cosh t
    {\partial \over \partial t }  + 2   \sinh t ) \right.
    \nonumber\\
   \left.  -
  ( { \partial^{2} \over \partial r^{2}}   + {2 \over \tan r} {\partial \over\partial r} )
  + {2\nu^{2} \over \sin^{2} r} \right ] g_{1}
  \nonumber\\
  =
  {\nu  F \over \sin^{2} r}     +  { 1 \over \sin r }( \partial _{r}  + {1 \over \tan r} ) F _{2}
   \, .
 \nonumber
 \end{eqnarray}

The right hand  part  of the equation turns to vanish due to the first relation
in  (\ref{A.56}). Thus, we arrive at the equation
\begin{eqnarray}
 \left [   \cosh t {\partial \over \partial t} (\cosh t     {\partial \over \partial t  }  + 2   \sinh t )  \right.
 \nonumber\\
\left. -
  ( { \partial^{2} \over \partial r^{2}}   + {2 \over \tan r} {\partial \over\partial r} )
  + {2\nu^{2} \over \sin^{2} r} \right ] g_{1}=0 \, ,
  \label{A.67a}
  \end{eqnarray}

\noindent
differently it reads as
\begin{eqnarray}
 \left [   { \partial ^{2} \over \partial t^{2}}  +  3 \tanh t {\partial \over \partial t}-{1 \over \cosh^{2} t}
 \left  (  { \partial ^{2} \over \partial r^{2} }
     \right.\right.
  \nonumber\\
  \left.\left.+ {2 \over \tan r } {\partial \over \partial r }- { j(j+1) \over \sin^{2} r } \right )  + 2\right  ] g_{1}=0  .
   \label{A.67b}
   \end{eqnarray}

Evidently, it is  $(t,r)$-part of the conformally invariant massless wave equation
in de Sitter space
\begin{eqnarray}
\left (
 {1 \over \sqrt{-g} } {\partial \over \partial x^{\alpha } } \sqrt{-g } g^{\alpha \beta } { \partial \over \partial x^{\beta}} +2 \right ) \Phi =0\,.
 \label{A.68}
 \end{eqnarray}

\noindent
Equation (\ref{A.67b}) with the help of substitution
\begin{eqnarray}
g_{1} = {g_{1}(t)  \over \cosh t} \; {g_{1}(r) \over \sin r}
\nonumber
\label{A.69a}
\end{eqnarray}

\noindent
leads to equations in separated variables
\begin{eqnarray}
\left ( {d^{2} \over dr^{2}} + \omega^{2} - {j(j+1) \over \sin^{2} r} \right ) g_{1}(r) = 0\,,
\label{A.69b}
\\
\left ( \cosh  {d \over d t} \cosh t {d \over dt } + \omega^{2} \right ) g_{1}(t) =0
\nonumber\\
\mbox{or} \qquad ({d ^{2} \over d \tau^{2}} + \omega^{2} ) g_{1} (\tau)=0\,.
\label{A.69c}
\end{eqnarray}

The functions  $g_{1}(t,r)$, determined by
\begin{eqnarray}
 {\partial \over \partial \tau } g_{2} = - {\nu \over \sin r}  g_{1} +{ F / 2 \over \sin r}  \, ,
 \nonumber\\
 {\partial \over \partial \tau } g_{3}= {\partial \over \partial r} g_{1}+= {F _{2} \over \sin r}  \, ,
\label{A.70}
\end{eqnarray}

\noindent
provide us with the complete description of electromagnetic waves of electric type in Lorentz gauge.

\section{Gradient type solutions
}

Among all electromagnetic solutions here exist solution of a pure gauge nature,
that are constructed as gradient of a scalar function; therefore having trivial (vanishing) electromagnetic tensor:
\begin{eqnarray}
f_{6} =0, \;\;
f_{7} =0, \;\; f_{8} =0, \;\; f_{9} =0, \;\; f_{10} =0 \, .
\nonumber
\end{eqnarray}

It is easily see that such solutions cannot have parity
 $ P = (-1)^{j+1}$
(see (\ref{A.46})):
\begin{eqnarray}
-\cosh t  {\partial \over \partial t} f_{5}  - 2\sinh t f_{5}
\nonumber\\
+ i(
{\partial \over \partial r}  + { 1 \over  \tan r} ) f_{8}+ { i \nu
/\sqrt{2} \over \sin r }  f_{9}=0 \, ,
\nonumber
\\
\cosh t ( { \partial \over \partial t} f_{2} -  f_{5} ) + \sinh t
f_{2} = 0 \, ,
\nonumber
\\
-  \cosh t f_{8} - i ( {\partial \over \partial r}   + { 1 \over
\tan r } )  f_{2}  =0 \, ,
\nonumber\\
 - \cosh t  f_{9}  + { i \sqrt{2}
\nu   \over \sin r }   f_{2}=0 \, , \label{A.71}
\end{eqnarray}

\noindent
as according to (\ref{A.65}) components  $f_{8}, f_{9}$ do not vanish.

For states with opposite parity (see (\ref{A.53a})  and  (\ref{A.53b}))

\vspace{2mm} $ P=(-1)^{j} $
\begin{eqnarray}
(  {\partial \over \partial r}    + {2 \over  \tan r}   )  f_{6}
+{ 2\nu  \over \sin r } \;
 f_{5}   =0 \, ,
\nonumber
\\
-\cosh t  ({\partial \over \partial t} + 2\tanh  t  ) f_{5}
\nonumber\\
+
 i(   {\partial \over \partial r}  + { 1 \over  \tan r} )  f_{8}    =0 \, ,
\nonumber
\\
-\cosh t (  {\partial \over \partial t}    +  2\tanh  t  ) f_{6}-
{ 2i  \nu  \over \sin r }  f_{8}    =0 \, , \label{A.72a}
\\[2mm]
\cosh t  ( { \partial \over \partial t} + \tanh t )  f_{2}
\nonumber\\
+
 { \nu  \over \sin r }  f_{1}  -   \cosh t f_{5} = 0 \, ,
\nonumber
\end{eqnarray}
\begin{eqnarray}
- \cosh t ( {\partial \over \partial t}  + \tanh  t)  f_{3}
\nonumber\\
  +
 {\partial \over \partial r} f_{1}   +   \cosh t f_{6} =0\, ,
\nonumber
\\
 i ({\partial \over \partial r}    + { 1 \over  \tan r} )  f_{2} +
{ i  \nu  \over \sin r }  f_{3} +  \cosh t  f_{8} =0 \,;
\label{A.72b}
\end{eqnarray}

\noindent
here one must set
$f_{5} =f_{6}=f_{8} =0;
$  and get three identities $0=0$; three reaming equations will take the form
\begin{eqnarray}
\cosh t  ( { \partial \over \partial t} + \tanh t )  f_{2} +
 { \nu  \over \sin r }  f_{1} = 0 \; ,
\nonumber
\\
- \cosh t ( {\partial \over \partial t}  + \tanh  t)  f_{3}   +
 {\partial \over \partial r} f_{1}    =0\; ,
\nonumber
\\
  ({\partial \over \partial r}    + { 1 \over  \tan r} )  f_{2} +
{   \nu  \over \sin r }  f_{3}  =0 \; .
\label{A.73}
\end{eqnarray}

\noindent
Translating the system to functions $g_{j}$:
\begin{eqnarray}
f_{1} = {g_{1}  \over \cosh t  } \, ,  \quad f_{2} = {g_{1}
\over \cosh t  } \, , \quad
 f_{3} = {g_{1}  \over \cosh t  } \, ,
\nonumber
\end{eqnarray}

\noindent
we have
\begin{eqnarray}
\cosh t   { \partial \over \partial t}  g_{2} +
 { \nu  \over \sin r }  g_{1} = 0 \, ,
 \nonumber\\
- \cosh t  {\partial \over \partial t} g_{3}   +
 {\partial \over \partial r} g_{1}    =0\, ,
\nonumber
\\
  ({\partial \over \partial r}    + { 1 \over  \tan r} )  g_{2} +
{   \nu  \over \sin r }  g_{3}  =0 \, .
\label{A.74a}
\end{eqnarray}

\noindent With the use of the variable  $\tau$, eqs.  (\ref{A.74a}) are written as
\begin{eqnarray}
{ \partial \over \partial \tau }  g_{2} +
 { \nu  \over \sin r }  g_{1} = 0 \, , \;\;
-   {\partial \over \partial \tau } g_{3}   +
 {\partial \over \partial r} g_{1}    =0\, ,
\nonumber
\\
  ({\partial \over \partial r}    + { 1 \over  \tan r} )  g_{2} +
{   \nu  \over \sin r }  g_{3}  =0 \, .
\label{A.74b}
\end{eqnarray}

\noindent
Note that the third equation is a consequence of two first.
As must be expected, these two first independent equations
coincide with
 (\ref{A.70})
\begin{eqnarray}
 {\partial \over \partial \tau } g_{2} + {\nu \over \sin r}  g_{1} ={ F / 2 \over \sin r}  \, ,
 \quad
 {\partial \over \partial \tau } g_{3} - \nu {\partial \over \partial r} g_{1} = F _{2}  \, ,
\nonumber
\end{eqnarray}

\noindent
at  $F_{2}=0, \; F=0$.

At considering electromagnetic field, as in Minkowski space
it is possible to impose Lorentz conditions
\begin{eqnarray}
  \nabla _{\alpha }  \Psi ^{\alpha } = 0 \, .
\label{A.75a}
\end{eqnarray}

\noindent
Let translate this relation to DKP basis. Instead of the 4-vector   $\Psi
^{\alpha }$   we should introduce its tetrad components
\begin{eqnarray}
\Psi ^{\alpha } = e^{(a)\alpha } \; \Psi _{a} \, ,
\nonumber
\end{eqnarray}
where
\begin{eqnarray}
e^{(0)\alpha} = ( 1 , 0 , 0 , 0 ) \, ,
\nonumber\\
 e^{(1) \alpha } = (
0 , 0 ,{ -1 \over   \cosh t \,\sin r}, 0 )\,  ,
\nonumber\\
e^{(2)\alpha } = ( 0 ,0 , 0 , {- 1 \over \cosh t \, \sin r \,  \sin
\theta})\, ,
\nonumber\\
 e^{(3)\alpha } = ( 0 , {- 1 \over   \cosh t }
, 0 , 0 ) \, .
\nonumber
\end{eqnarray}

\noindent
Correspondingly,   (\ref{A.75a}) takes the form
\begin{eqnarray}
 e^{(b)\alpha }_{;\alpha } \; \Psi _{b}
\; +  \; e^{(b)\alpha } \; \partial _{\alpha }  \Psi _{b}
  =0    \, .
\label{A.75c}
\end{eqnarray}

\noindent We will need  the formulas
\begin{eqnarray}
e^{(0)\alpha }_{;\alpha }
   = 3  \tanh t \,, \;\;\;  e^{(3)\alpha }_{;\alpha }=-
{2 \over \cosh t}{1 \over \tan r}\,,
\nonumber\\
e^{(1)\alpha }_{;\alpha } =- {1 \over \cosh t} {1 \over \sin r}
{\cos \theta \over \sin \theta} \, , \;\;\; e^{(2)\alpha
}_{;\alpha }=
  0\,.
\nonumber
\end{eqnarray}

\noindent Tetrad components of   $\Psi _{a}$ are referred to the first four components
of DKP function in cyclic basis
by the formulas  (let $W \equiv 1/ \sqrt{2}$)
\begin{eqnarray}
\left | \begin{array}{l}
\Phi_{0} \\ \Phi_{1} \\ \Phi_{2} \\ \Phi_{3} \\
        \end{array} \right | =
\left | \begin{array}{rrrr}
1      & 0 & 0 &  0     \\
0      &-W & 0 & +W      \\
0      &-iW& 0 &-iW       \\
0      & 0 & 1 &  0
\end{array} \right | =
\left | \begin{array}{l}
    f_{1} \;D_{0}    \\  f_{2} \;D_{-1} \\  f_{3} \;D_{0} \\
    f_{4} \; D_{+1}
\end{array} \right |,
\nonumber
\end{eqnarray}

\noindent so we have
\begin{eqnarray}
\Psi _{0} =  f_{1} \; D_{0 } \, , \quad \Psi _{3} =  f_{3} \;
D_{0 } \, ,
\nonumber
\\
\Psi _{1} = {1 \over \sqrt{2}}  (- f_{2} D_{ -1} + f_{4}  D_{ +1}
) \, ,\nonumber
\\
\Psi _{2} =  {i \over \sqrt{2}}  (- f_{2} D_{ -1}   - f_{4}  D_{
+1} ) \, . \label{A.76}
\end{eqnarray}

\noindent
Eq. (\ref{A.75c}) can be written as
\begin{eqnarray}
 e^{(0)\alpha }_{;\alpha } \; \Psi _{0} +
  e^{(3)\alpha }_{;\alpha } \; \Psi _{3}+
   e^{(1)\alpha }_{;\alpha } \; \Psi _{1}+
    e^{(2)\alpha }_{;\alpha } \; \Psi _{2}
\nonumber\\
 +   e^{(0)\alpha } \; \partial _{\alpha }  \Psi _{0}
 +   e^{(3)\alpha } \; \partial _{\alpha }  \Psi _{3}
 \nonumber\\
 +   e^{(1)\alpha } \; \partial _{\alpha }  \Psi _{1}
 +   e^{(2)\alpha } \; \partial _{\alpha }  \Psi _{2}
  =0    \, ,
\nonumber
\end{eqnarray}

\noindent which is equivalent to
\begin{eqnarray}
3  \tanh t\; \Psi _{0}
  - {2 \over \cosh t}{1 \over \tan r}\; \Psi _{3}
   \nonumber\\
   - {1 \over \cosh t} {1 \over \sin r} {\cos \theta \over \sin \theta}  \; \Psi _{1}
 +   \partial_{t}   \Psi _{0}
 -{1 \over \cosh t} \partial _{r }  \Psi _{3}
 \nonumber\\
 - { -1 \over   \cosh t \sin r} \partial _{\theta }  \Psi _{1}
 -   { 1 \over \cosh t  \sin r \;  \sin \theta}  \partial _{\phi }  \Psi _{2}
  =0    \, ,
\nonumber
\end{eqnarray}

\noindent and further with the use of (\ref{A.70}) it reads
 \begin{eqnarray}
3  \tanh t  f_{1}  D_{0 }
  - {2 \over \cosh t}{1 \over \tan r}\; f_{3}  D_{0 }
   \nonumber\\
   - {1 \over \cosh t} {1 \over \sin r} {\cos \theta \over \sin \theta}
   {1 \over \sqrt{2}}  (- f_{2} D_{ -1} + f_{4}  D_{ +1} )
   \nonumber\\
 +   \partial_{t}   f_{1}  D_{0 }
 -{1 \over \cosh t} \partial _{r } f_{3} D_{0 }
 \nonumber\\
 - { 1 \over   \cosh t \sin r} \partial _{\theta }  {1 \over \sqrt{2}}  (- f_{2} D_{ -1} + f_{4}  D_{ +1} )
 \nonumber\\
  -   { 1 \over \cosh t  \sin r \,  \sin \theta}
   \partial _{\phi }  {i \over \sqrt{2}}  (- f_{2} D_{ -1}   - f_{4}  D_{ +1} )
  =0     .
\nonumber
\end{eqnarray}

\noindent
The last equation, after regrouping and some elementary manipulation
takes the form
 \begin{eqnarray}
  {1 \over \sin r}{1 \over \sqrt{2}}\left [      f_{2} \left (  {\cos \theta \over \sin \theta}
   +       \partial _{\theta }
  -  { m \over     \sin \theta} \right )  D_{ -1}  \right.
  \nonumber\\
   \left. -
   f_{4} \left ( {\cos \theta \over \sin \theta}  +  \partial _{\theta }   +
  { m \over     \sin \theta}  \right ) D_{ +1} \right ]
\nonumber
\\
  +
   \left [   \cosh t  (\partial_{t}
 +3  \tanh t ) f_{1} \right.
 \nonumber\\
 \left.- ( \partial _{r}
  + {2 \over \tan r} ) f_{3}
      \right ] D_{0 }
=0    \, .
\label{A.77}
\end{eqnarray}

\noindent
 Finally, with the use of recurrent relations for Wigner functions
 \begin{eqnarray}
\left ( \partial _{\theta }  - {m  - \cos \theta \over \sin \theta
} \right ) D_{ -1} = - \sqrt{(j  +1)j } \; D_{0 } \,,
\nonumber\\
\left ( \partial _{\theta } + {m + \cos \theta \over \sin \theta }
\right )  D_{ +1} = - \sqrt{(j +1)j} \; D_{0 }    \, ,
\nonumber
\end{eqnarray}

\noindent we derive  the following equation
(remember that $\sqrt{j(j+1)/2} =
\nu $)
\begin{eqnarray}
  - { \nu   \over \sin r} (     f_{2}
    +
   f_{4} )  +     \cosh t   (\partial_{t}  +3  \tanh t ) f_{1}
   \nonumber\\
    - ( \partial _{r}
  + {2 \over \tan r} ) f_{3}
=0    \, . \label{A.78}
\end{eqnarray}

\noindent
This is the Lorentz condition after excluding the $(\theta, \phi)$ dependence.

For states with the parity
\begin{eqnarray}
P = (-1)^{j+1} , \quad f_{1} = \; f_{3} =  0 \, , \quad f_{4}= -
f_{2} \, ,
\nonumber
\end{eqnarray}
it becomes identity $  0 = 0$. For states with the opposite parity we have
\begin{eqnarray}
P = (-1)^{j}, \quad  f_{4} = + \; f_{2} \, ,
\nonumber
\\
 - { 2\nu   \over \sin r}      f_{2}     +
  \cosh t     (\partial_{t}  +3  \tanh t ) f_{1}
  \nonumber\\
  - ( \partial _{r}
  + {2 \over \tan r} ) f_{3}
=0    \, .
\label{A.80}
\end{eqnarray}

Translating it to the functions
\begin{eqnarray}
f_{1} = {g_{1}  \over \cosh t  } \, ,  \quad f_{2} = {g_{1}
\over \cosh t  } \, , \quad
 f_{3} = {g_{1}  \over \cosh t  } \, ,
\nonumber
\end{eqnarray}

\noindent
we get
\begin{eqnarray}
 - { 2\nu   \over \sin r}      g_{2}     +
  (\cosh t     \partial_{t}  + 2   \sinh t ) g_{1}
  \nonumber\\
  - ( \partial _{r}
  + {2 \over \tan r} ) g_{3}
=0    \, .
\label{A.81}
\end{eqnarray}

Let  the above gradient type solution
\begin{eqnarray}
- \cosh t   { \partial \over \partial t}  g_{2}  =
 { \nu  \over \sin r }  g_{1}  \, ,
 \nonumber\\
 \cosh t  {\partial \over \partial t} g_{3} =  {\partial \over \partial r} g_{1}    \, ,
\label{A.82}
\end{eqnarray}

\noindent obey the Lorentz condition. To take into account (\ref{A.82}) in Lorentz condition (\ref{A.81}), one should
act on eq.  (\ref{A.81}) by the operator
$
\cosh t   { \partial \over \partial t}$;
 after that we derive equation for  $f_{1}$
\begin{eqnarray}
\left [
   \left (  { \partial^{2}  \over \partial t^{2} } + 3  \tanh t {\partial \over \partial t}
+ 2
\right )-{1 \over \cosh^{2} t }  \left (
  {\partial^{2} \over \partial r^{2} } \right.\right.
\nonumber\\
\left. \left.+
 {2  \over \tan r } {\partial \over \partial r}  - { 2\nu^{2} \over \sin^{2} r } \right ) \right ] g_{1}=0\,.
\label{A.83a}
\end{eqnarray}
Evidently, this is a  $(t,r)$-part of Klein--Fock--Gordon equation with conformal term
\begin{eqnarray}
\left ( {1 \over \sqrt{-g} } \partial_{\alpha} \sqrt{-g} g^{\alpha \beta} \partial_{2} + 2 \right )
\Phi =0 \,.
\label{A.83b}
\end{eqnarray}
Eq.  (\ref{A.83a}) is solved  trough separation of the variables (see  (\ref{A.69b}), (\ref{A.69c}), and then one can obtain expressions for
$g_{2},\, g_{3}$ -- see  (\ref{A.76})).

\section{Conclusion}

Tetrad-based complex  formalism by
Majorana--Oppenheimer has been applied  to
treat electromagnetic field in extending de Sitter Universe in
non-static spherically-symmetric coordinates. With the help of
Wigner $D$-function we separate angular dependence in the complex
vector field $E_{j}(t,r)+i B_{j}(t,r)$ from $(t,r)$-dependence.
After that we separate variables $t$ and $r$.
Non-static  geometry of the de Sitter model lead to definite dependence of electromagnetic modes
 on the time variable.
Relation of 3-vector complex approach to 10-dimensional Diffin--Kemmer-Petiau formalism is
elaborated. On this base,   the electromagnetic wave of magnetic and
electric type have been constructed in both approaches, also
in DK form, there are specified gradient-type solutions in Lorentz gauge.

This  work was   supported   by the Fund for Basic Researches of Belarus,
 F 13K-079, within the cooperation framework between Belarus  and Ukraine.

N.~D.~Vlasii and Yu.~A.~Sitenko were supported by the State
Agency for Science, Innovations and Informatization of Ukraine
under the SFFR-BRFFR grant F54.1/019.

\end{document}